\documentclass{sig-alternate}

\newcommand{\ignore}[1]{}
\usepackage{fancyhdr}
\usepackage[normalem]{ulem}
\usepackage[hyphens]{url}
\usepackage[bookmarks=true,breaklinks=true,letterpaper=true,colorlinks,linkcolor=black,citecolor=blue,urlcolor=black]{hyperref}
\usepackage{xspace}
\usepackage{color}
\usepackage[table]{xcolor}
\usepackage[capitalize]{cleveref}
\usepackage{paralist}
\usepackage[linesnumbered, ruled]{algorithm2e}
\usepackage{algorithmic}
\usepackage{subfig}
\usepackage{soul}
\usepackage{comment}
\usepackage{times}

\newcommand{\name}{Banshee\xspace}
\definecolor{comment-color}{rgb}{0.25,0.25,0.25}
\newcommand{\codeComment}[1]{\textnormal{\color{comment-color}{\textit{\textbf{\# 
#1}}}}\unskip}
\SetKwRepeat{Do}{do}{while}
\soulregister\cite7
\soulregister\ref7
\soulregister\texttt7
\soulregister\textbf7
\soulregister\textit7
\soulregister\cref7
\soulregister\name7

\newcommand{\rev}[1]{{#1}}

\title{Banshee: Bandwidth-Efficient DRAM Caching Via Software/Hardware Cooperation}
\begin{document}

\author{
\vspace{-.6in} \\
\normalsize Xiangyao Yu$^\dagger$, Christopher J.  Hughes$^\ddagger$, 
Nadathur Satish$^\ddagger$, Onur Mutlu$^\mathsection$, Srinivas 
Devadas$^\dagger$ \vspace{.05in}
\\
\normalsize $^\dagger$ CSAIL, MIT, $^\ddagger$ Intel Labs,  
$^\mathsection$ ETH Zurich \vspace{.05in} \\
\normalsize
$^\dagger\{$yxy, devadas$\}$@mit.edu,
$^\ddagger\{$christopher.j.hughes, 
nadathur.rajagopalan.satish$\}$@intel.com \\
\normalsize 
$^\mathsection$onur.mutlu@inf.ethz.ch
}

\maketitle
\pagestyle{plain}

\begin{abstract}

Putting the DRAM on the same package with a processor enables several
times higher memory bandwidth than conventional off-package DRAM. Yet,
the latency of in-package DRAM is not appreciably lower than that of
off-package DRAM. A promising use of in-package DRAM is as a large
cache. Unfortunately, most previous DRAM cache designs mainly optimize
for hit latency and do not consider off-chip bandwidth efficiency as a
first-class design constraint. Hence, as we show in this paper, these
designs are suboptimal for use with in-package DRAM.

%%% ONUR: We should be clear about which bandwidth we are talking
%%% about when we say we want to improve bandwidth efficiency. Do we
%%% want to improve the wasted bandwidth of the DRAM cache or the
%%% wasted bandwidth of off-chip memory?

We propose a new DRAM cache design, \name, that optimizes for both in-
and off-package DRAM bandwidth efficiency without degrading access
latency.  The key ideas are to eliminate the in-package DRAM bandwidth 
overheads due to costly tag accesses through virtual memory mechanism 
and to incorporate a bandwidth-aware frequency-based replacement 
policy that is biased to
reduce unnecessary traffic to off-package DRAM. Our extensive
evaluation shows that \name provides significant performance
improvement and traffic reduction over state-of-the-art
latency-optimized DRAM cache designs.

%Vendors are rolling out microprocessors that couple high-capacity
%DRAMs and high-performance, throughput-oriented processors. Putting
%the DRAM into the same package with a processor allows for several
%times higher memory bandwidth than conventional off-package DRAMs.
%The latency of this new memory, however, is \textbf{not} appreciably
%lower than external DDR.

%Processor designers would like to enable software to automatically see
%benefits from this new memory by using it as a cache of external
%memory. Most previous DRAM cache designs, however, mainly optimized
%for hit latency and failed to achieve good bandwidth efficiency.  This
%leads to suboptimal performance for memory intensive applications.

%We propose \name, a bandwidth-optimized DRAM cache design for systems
%with high bandwidth in-package DRAM. It largely eliminates bandwidth
%overheads from tag accesses by storing the mapping information in the
%page table, and uses a bandwidth-aware frequency-based replacement
%policy biased against replacement to reduce unnecessary DRAM traffic.
%Our simulation shows that \name can provides significant performance
%improvement and DRAM traffic reduction over state-of-the-art
%latency-optimized designs.

\end{abstract}

\section{Introduction} \label{sec:intro}

In-package DRAM technology
integrates the CPU and a high-capacity multi-GB DRAM in the same 
package, enabling much higher bandwidth than traditional off-package 
DRAM. For emerging memory bandwidth-bound applications (e.g., graph 
and machine learning algorithms, sparse linear algebra-based HPC 
codes), in-package DRAM can significantly boost system 
performance~\cite{hbm2014, hbm-amd}.  Several hardware vendors are 
either offering or will soon offer processors with in-package DRAM 
(e.g., Intel's Knights Landing~\cite{knl-micro}, AMD's 
Fiji~\cite{amdfuji2015}, and Nvidia's Pascal~\cite{pascal2014}) and a 
large number of designs have been proposed in both industry and 
academia~\cite{loh2011, qureshi2012, jevdjic2014, chou2015, lee2015, 
sim2014, chou2014}.

% CJH: for references above...
% KNL link: https://software.intel.com/en-us/blogs/2016/01/20/an-intro-to-mcdram-high-bandwidth-memory-on-knights-landing
% wikipedia's page on HBM: https://en.wikipedia.org/wiki/High_Bandwidth_Memory
%   The latter mentions these initial products: AMD's Fiji and Arctic Islands, Nvidia's Pascal

% TODO is there a fundamental reason why in-package drams are not 
% improveing latency?

One critical property of in-package DRAM is that, while it
provides high bandwidth, \textit{its latency will still be similar to 
or even worse than off-package DRAM}~\cite{knl2015, hmc2.1}. This is 
one of the reasons why the products first incorporating it are all in 
the throughput computing space, where the target applications are 
typically latency-tolerant, but very bandwidth-hungry. Many previous 
DRAM cache designs, however, assumed low latency in-package DRAM and 
threfore are not necessarily the best fit.  

%assumed in-package DRAM has lower latency and thus optimized for 
%latency.  Many previous works proposed using in-package DRAM as a
%hardware-managed cache~\cite{loh2011, qureshi2012, jevdjic2014}; most 
%assume that in-package DRAM will provide lower latency than 
%off-package memory, and thus optimize
%for latency. %and evaluate their design for latency-bound workloads.

%Since this assumption is not true for at least the initialproducts, 
%previous designs are not necessarily the best fit.

In particular, many of the designs incur large amounts of traffic to 
in-package and/or off-package DRAM for \textit{meta data management} 
(e.g., tags, LRU bits) and \textit{cache replacement}. 
%Since the tag array of a multi-gigabyte cache cannot reasonably fit 
%in on-die SRAM, many previous designs (especially cacheline 
%granularity designs) store the tag array in in-package DRAM 
%itself~\cite{qureshi2012, jevdjic2014} which is accessed for each tag 
%lookup.  Although the latency of a tag lookup can be largely hidden 
%using previously proposed techniques, the extra data accessed 
%consumes valuable bandwidth.
In page-granularity DRAM caches, previous works (e.g., Tagless DRAM 
cache, TDC~\cite{lee2015, jang2016}) have proposed storing the page 
mapping information in the Page Table Entries (PTEs) and Translation 
Lookaside Buffers (TLBs), by giving different physical address regions 
to in- and off-package DRAMs. This completely removes the bandwidth
overhead for tag lookups.  However, the bandwidth inefficiency for 
\textit{DRAM cache replacement} still remains.
%since frequently replacing data at page granularity can be expensive.  
Some techniques have been proposed to improve replacement bandwidth 
efficiency (e.g., footprint cache~\cite{jang2016, jevdjic2013} and 
frequency based replacement~\cite{jiang2010}), but existing solutions 
still incur significant overhead. 

Supporting efficient replacement in PTE/TLB-based DRAM cache designs 
is inherently difficult due to the 
%\textit{address consistency}  and 
\textit{TLB coherence} problem. 
%Since moving a page in/out of the DRAM cache changes its physical 
%address, all of the on-chip caches must be scrubbed of cachelines on 
%the remapped page for address consistency, which incurs significant 
%overhead on every replacement.  The address consistency problem was 
%not fully addressed in TDC.  Further, 
When a page is remapped, an expensive mechanism is required to keep 
all TLBs coherent. Due to the complexity, previous work had certain 
requirements with respect to when replacement is allowed to happen 
(e.g., on every miss~\cite{lee2015}) making it hard to design 
bandwidth efficient replacement.

In this paper, we propose \textit{\name}, a DRAM cache design aimed at
maximizing the bandwidth efficiency of both in- and off-package DRAM,
while also providing low access latency. Similar to
TDC~\cite{lee2015}, \name avoids tag lookup by storing DRAM cache 
presence information in
the page table and TLBs. \name's key innovation
over TDC is its bandwidth-efficient replacement policy, and design
decisions that enable its usage.  Specifically,
%Similar to previous work~\cite{jiang2010}, 
\name uses a hardware-managed frequency-based replacement (FBR) policy
that only caches hot pages to reduce unnecessary data replacement
traffic. To reduce the cost of accessing/updating frequency counters
(which are stored in in-package DRAM), \name uses a new
\textit{sampling} approach to only read/write counters for a fraction
of memory accesses. Since \name manages data at page granularity,
sampling has minimal effect on the accuracy of frequency prediction.
This strategy significantly brings down the bandwidth overhead of
cache replacement. The new replacement policy also allows \name to 
support large (2~MB) pages efficiently with simple extensions.  
Traditional page based DRAM cache algorithms, in contrast, failed to 
cache large pages due to the overhead of frequent page 
replacement~\cite{jang2016}.  

To enable the usage of this replacement scheme, we need new techniques
to simplify TLB coherence. \name achieves this by not updating the 
page table and TLBs for
every page replacement, but only doing so lazily in batches to 
amortize the cost. The batch update mechanism is implemented
through software/hardware co-design where a small hardware table (Tag 
Buffer) maintains the up-to-date mapping information at
each memory controller, and triggers the software routine to update
page tables and TLBs whenever the buffer is full.  

Specifically, \name makes the following contributions:

\begin{compactenum}
    
    \item \rev{\mbox{\name} significantly improves the bandwidth 
    efficiency for DRAM cache replacement through a bandwidth-aware 
    frequency-based replacement policy implemented in hardware. It 
    minimizes unnecessary data and meta-data movement. } 
    
    \item \rev{\mbox{\name} resolves the address consistency problem,
      and greatly simplifies the TLB coherence problem that are in
      previous PTE/TLB-based DRAM cache designs, via a new,
      \textit{lazy} TLB coherence mechanism. This allows more
      efficient replacement policies to be implemented. }
   
    \item By combining PTE/TLB-based page mapping management and
	efficient hardware replacement, \name significantly improves 
	in-package DRAM bandwidth efficiency.  
	%, \mbox{\name} achieves 	near-optimal in-package DRAM bandwidth 
	%efficiency. 
	Compared to other three state-of-the-art DRAM cache designs, \name 
	outperforms the best of them (Alloy Cache~\cite{qureshi2012}) by 
	15.0\% and reduces in-package DRAM traffic by
    35.8\%. 
	
	\item \name can efficiently support large pages (2~MB) using 
	PTEs/TLBs. Replacement overhead of large pages is significantly 
	reduced through our bandwidth-efficient replacement policy.  

%%% ONUR: Do we show near-optimalilty as we claim above? Can we add a forward pointer to that section?
%%% ONUR: what is the best previous scheme we evaluated. Can we cite here?

\end{compactenum}

\section{Background} \label{sec:back}

In this section, we discuss the design space of DRAM caches,
and where previous proposals fit in that space.  We focus on
two major considerations in DRAM cache designs: how to track the
contents of the cache (\cref{sec:tracking}), and how to change the 
contents (i.e.,
replacement, \cref{sec:background-replacement}). 
%Specifically, we will show how previous works failed to efficiently 
%utilize the precious DRAM bandwidth.   

%Previous work has considered a variety of policies for both of these, 
%as well as different granularities of data to store in the cache, and 
%we will cover this as well; however, this decision is often tightly 
%coupled to the tracking and placement policies.

%For our discussion, we make some assumptions about the architecture
%of the memory hierarchy.  First, we assume the processor has
%an SRAM last-level cache (LLC), with a line size of 64B, and
%an address size of 48 bits.
%The system does \textit{not} necessarily maintain inclusion
%between the LLC
%and the in-package DRAM.  Misses in the LLC go to a set
%of memory controllers shared by all cores --- the physical
%address space is partitioned amongst the controllers according
%to some hashing function, to avoid hotspots.
%We assume this hashing is done at 4KB page granularity.
%%, i.e., each 4KB page maps to a single memory controller.
%The memory controllers each are responsible for one or more
%channels of in- and off-package DRAM.
%We assume that memory controllers do not send each other requests;
%thus, the physical address of a line determines which portion of
%in-package DRAM the line can reside in.

For our discussion, %Regarding the physical design of the in-package 
%DRAM,
we assume the processor has an SRAM last-level cache (LLC) managed at 
cacheline (64~B) granularity.  Physical
addresses are mapped to memory controllers (MC) statically at page 
granularity (4~KB).
We also assume the in-package DRAM is similar to the first-generation 
High Bandwidth Memory (HBM).  The link width between the memory 
controller and HBM is 16B, but with a minimum data transfer size of 
32B~\cite{hbm2014}. Thus, reading a 64B cache line plus the tag 
transfers at minimum 96B. We also assume the in- and off-package DRAMs 
have the same latency.

\begin{table*}
    \caption{ Behavior of different DRAM cache designs.  Assumes
    perfect way prediction for Unison Cache.
    Latency is relative to access time for off-package DRAM.}
    \vspace{-.2in}
    \begin{center}
    {
		\footnotesize
	\begin{tabular}{|c|p{1.45in}|p{1.35in}||p{.9in}|p{1.15in}|p{0.65in}|}
    \hline
	Scheme & DRAM Cache Hit & DRAM Cache Miss & Replacement Traffic & 
	Replacement \newline Decision & Large Page Caching
	\\ \hline
    \hline
    % Here, I assume UnisonCache needs to update LRU bits for each 
    % access. That's why its 128 B. 
    Unison & \cellcolor[rgb]{.95,.8,.8} Traffic: At least 128B 
    \newline
           (data + tag read/update) \newline Latency: $\sim$1x
           & \cellcolor[rgb]{.95,.8,.8} Traffic: At least 96B \newline
           (spec. data + tag read)
           %\newline footprint size + 32B for fill.  
           \newline Latency: $\sim$2x
           & \cellcolor[rgb]{.95,.8,.8} On every miss \newline
		   32B tag + Footprint size & \cellcolor[rgb]{.8, .9, .8}
		   Hardware managed, \newline way-associative, \newline LRU
		   & \cellcolor[rgb]{.95,.8,.8}No
    \\ \hline
    Alloy  & \cellcolor[rgb]{.95,.8,.8} Traffic: 96B \newline(data + 
    tag read) \newline Latency: $\sim$1x
           & \cellcolor[rgb]{.95,.8,.8} Traffic: 96B \newline
           (spec. data + tag read)
           %tag+data on lookup and fill).
           \newline Latency: $\sim$2x
           & \cellcolor[rgb]{.9,.9,.9} On some misses\newline
           32B tag + 64B fill
		   & \cellcolor[rgb]{.8,.9,.8} Hardware managed,\newline 
		   direct-mapped, \newline
		   stochastic~\cite{chou2015}
		   & \cellcolor[rgb]{.8,.9,.8}Yes
    \\ \hline
    TDC    & \cellcolor[rgb]{0.9,0.9,0.9} Traffic: 64B. \newline
           Latency: $\sim$1x \newline TLB coherence
           & \cellcolor[rgb]{0.9,0.9,0.9} Traffic: 64B. \newline
           Latency: $\sim$1x \newline TLB coherence
           & \cellcolor[rgb]{0.95,0.8,0.8} On every miss \newline 
           Footprint size~\cite{jang2016}
		   & \cellcolor[rgb]{0.8,0.9,0.8} Hardware managed,\newline
		   fully-associative, \newline FIFO
		   & \cellcolor[rgb]{.95,.8,.8}No
    \\ \hline
    HMA    & \cellcolor[rgb]{0.8, 0.9, 0.8} Traffic: 64B. \newline 
           Latency: $\sim$1x
           & \cellcolor[rgb]{0.8, 0.9, 0.8} Traffic: 0B.  \newline 
           Latency: $\sim$1x
		   & \multicolumn{2}{p{2.21in}|}{
           \cellcolor[rgb]{0.95, 0.8, 0.8} Software managed,
		   high replacement cost} 
		   & \cellcolor[rgb]{.8,.9,.8}Yes
           %& \cellcolor[rgb]{0.95, 0.8, 0.8} Software managed, low 
           %traffic
           %& \cellcolor[rgb]{.95,.8,.8} High miss rate
    \\ \hline
    \name  & \cellcolor[rgb]{0.8, 0.9, 0.8} Traffic: 64B. \newline 
           Latency: $\sim$1x
           & \cellcolor[rgb]{0.8, 0.9, 0.8} Traffic: 0B
           \newline Latency: $\sim$1x
           & \cellcolor[rgb]{0.8, 0.9, 0.8} Only for hot pages\newline
           32B tag + page size
		   & \cellcolor[rgb]{.8, .9, .8} Hardware managed, \newline
		   way-associative, \newline frequency based
		   & \cellcolor[rgb]{.8,.9,.8}Yes
    \\ \hline
    \end{tabular}
    }
    \vspace{-.25in}
    \end{center}
    \label{tab:summary}
\end{table*}

\subsection{Tracking DRAM Cache Contents} \label{sec:tracking}

%\begin{figure}[t!]
%    \centering
%    \includegraphics[width=\columnwidth]{figs/motivation-hw.pdf}
%    \caption{ Overall Architecture }
%    \label{fig:arch}
%\end{figure}

For each LLC miss, % (or external coherence request),
the memory
controller determines whether to access the in-package or off-package DRAM.
Therefore, the mapping of each data block
% an index of the in-package DRAM cache contents must be 
must be stored somewhere in the system.

%To make this determination, it must track the contents of the 
%in-package DRAM cache. This tracking is also required for when we 
%replace a piece of data in the DRAM cache --- we must know what we're 
%replacing, at least if the data is dirty and must be written back to 
%a lower level of the memory hierarchy.

\subsubsection{Using Tags}

The most common technique for tracking the contents of a cache
is explicitly storing the tags for cached data. However, the tag 
storage can be significant when the DRAM cache is large. A 16~GB DRAM 
cache, for example, requires 512~MB (or 8~MB) tag storage if managed 
at cacheline (or page) granularity. %Since this is huge overhead for 
%Due to the large overhead, 
As a result, state-of-the-art DRAM cache designs store tags in the 
in-package DRAM itself. These designs, however, has the bandwidth 
overhead of tag lookup for each DRAM cache access.

\cref{tab:summary} summarizes the behavior
for some state-of-the-art DRAM cache designs, including two that
store tags in the in-package DRAM, \textit{Alloy Cache}~\cite{qureshi2012}
and \textit{Unison Cache}~\cite{jevdjic2014}.
%The second and third columns summarize the bandwidth and latency
%behavior for cache hits and misses, which we focus on here.
%For simplicity, %in the table and discussion,  we assume the in- and 
%off-package DRAM latencies are the same.

\textit{Alloy Cache} is a direct-mapped DRAM cache storing data in 
cacheline granularity. The tag and data for
a set are stored adjacently.
%and always accessed at the same time. 
On a hit, data and tag are read together with latency roughly that of 
a single DRAM access. On a miss, we pay the cost of a
hit plus the access to off-package DRAM and filling the data into
the DRAM cache. Therefore, both latency and bandwidth consumption may 
double. 
%--- thus, latency is 2$\times$ a DRAM access, and the
%DRAM cache bandwidth used is 2$\times$ the cost of a block + the tag.  
The original paper proposed to issue requests to in- and off-package 
DRAMs in parallel to hide miss latency. We disable this optimization 
here since it hurts performance when off-package DRAM bandwidth is 
scarce.   

%the latency is roughly that of a single
%DRAM access.  We must transfer both the block (64B), and the
%tag.  
\begin{comment}
The tag itself is small, but depending on the in-package
DRAM design, it may still consume significant bandwidth (e.g.,
32B, or 50\% overhead, for HBM).
\end{comment}

\textit{Unison Cache}~\cite{jevdjic2014} stores data in page 
granularity and supports set associativity. 
%is a design for storing pages rather than cachelines.  Unison Cache 
%supports set associativity --- it stores all of a set's tags 
%together, followed by the set's pages.
%To provide fast hit latency % without a huge bandwidth penalty from
%reading all of the data in the set, 
The design relies on way prediction to provide fast hit latency. On an 
access, the memory controller reads all of
the tags for a set plus the data only from the \textit{predicted} 
way. On a hit and correct way prediction, the latency is roughly
that of a single DRAM access; the data and tags are loaded and the LRU 
bits are updated.  
%on a misprediction, it will have to re-access the cache.
%On a hit and correct way prediction, the latency is roughly
%that of a single
%DRAM access.  We must transfer both the predicted block (64B), and 
%the tags ($\sim$32B for a 4-way set-associative cache) and update the 
%LRU bits. 
On a miss, latency is doubled, and we need extra 
traffic for off-package DRAM accesses and maybe cache replacement.
%to perform a tag lookup plus one block access (for speculative load) 
%and fill the page if replacement happens.
%potentially fill  plus a full page access (for the fill).

%To mitigate the large fill cost of reading a whole page, Unison Cache
%uses a sector cache design~\cite{rothman00}.  The cache requests data
%from off-package DRAM at the granularity of one block (64B), and for 
%each page in the cache, it tracks each individual block's presence.
%Unison Cache relies on a ``footprint predictor''~\cite{kumar1998} to 
%say which blocks to bring in on a cache miss.

\subsubsection{Using Address Remapping} \label{sec:remapping}

Another technique for tracking data in the DRAM cache is
%implicitly, 
via the virtual-to-physical address mapping~\cite{lee2015, 
meswani2015} in the page tables and TLBs. In these designs, the 
physical address space is carved up between in- and off-package DRAMs.  
Where a page is mapped to can be strictly determined using its 
physical address and the tag lookup is no longer required. 
%Since the TLB is accessed for each memory operation, it is virtually 
%free to determine mapping.  where a page is mapped to. 

Besides the TLB coherence challenge mentioned in \cref{sec:intro},  
%TLB coherence is a big challenge in these designs. A replacement 
%changes one or more pages' physical address.  We must ensure the 
%mappings in all cores' TLBs are kept coherent. Previous works have 
%proposed both software and hardware techniques for  this.
TLB/PTE-based designs have another challenge that we call 
\textit{address consistency}. When a page is remapped, its physical 
address is changed. Therefore, all of the on-chip caches must be 
scrubbed of cachelines on the remapped page to ensure consistent 
physical addresses. This leads to significant overhead for each page 
remapping.  

Heterogeneous Memory Architecture (HMA~\cite{meswani2015}) uses a 
software based solution to these problems.  
%uses address remapping, but relies on infrequent changes to the 
%cache's contents to avoid the above problems.
Periodically, the operating system (OS) ranks all pages and moves
hot pages into the in-package DRAM (and cold pages out). The OS
updates all PTEs, flushes all TLBs for coherence, and flushes remapped
pages from caches for address consistency. Due to the high cost, 
remapping can only be done 
%Such cache replacement is quite expensive---HMA does this at a very 
at a very coarse granularity (100~ms to 1 s) in order to amortize the 
cost.  Therefore, the replacement policy is not able to capture 
fine-grained temporal locality in applications.  Also, all programs 
running in the system have to stop when the pages are moved,
%between in- and off-package DRAMs
causing undesirable performance hiccups.

%\textbf{Hardware approach} (TDC~\cite{lee2015})

Tagless DRAM Cache (TDC~\cite{lee2015}) also uses address
remapping, but enables
frequent cache replacement via hardware-managed TLB coherence.
Specifically, TDC maintains a directory structure
in main memory and updates it whenever an entry is inserted 
or removed from any TLB. Such fine-grained TLB coherence incurs extra 
design complexity.  
%may be acceptable in terms of performance, since the latency of 
%accessing the directory can be hidden by the page table walking. The 
%complexity of fine-grained TLB coherence is still quite significant.  
Further, the storage of the directory may be a potential scalability 
bottleneck as core count increases.  The paper \cite{lee2015} does not 
discuss \textit{address consistency}, so
it is unclear which solution, if any, TDC employs.

\subsection{DRAM Cache Replacement}
\label{sec:background-replacement}

Cache replacement is another big challenge in in-package DRAM designs.  
%Since we cannot place all data in the in-package DRAM, we
%must use some replacement policy to determine which data (e.g., 
%blocks or pages) should be present at each point in time. 
We discuss both hardware and software approaches presented in previous 
work. %s here.  

\subsubsection{Hardware-Managed}

Hardware-managed caches are able to make placement decisions on each 
DRAM cache miss, and thus can adapt
rapidly to changing workload behavior.
%On a miss, hardware first decides \textit{whether} to insert
%the data in the DRAM cache.  
Many designs, including Alloy Cache, Unison Cache and TDC, always 
place the data in
the DRAM cache for each cache miss. Although this is common 
practice for SRAM caches, the incurred extra replacement traffic is 
quite expensive for DRAM.
Some previous designs try to reduce replacement traffic with a 
stochastic mechanism~\cite{chou2015} where replacement happens with a 
small probability at each access.  
For page-granularity DRAM cache designs, frequent replacement also 
causes \textit{over fetching}, where a whole page is cached but only a 
subset is actually accessed before eviction. For this problem, 
previous works proposed to use a sector cache design~\cite{rothman00} 
and rely on a ``footprint predictor''~\cite{kumar1998, jang2016} to 
determine which blocks to load on a cache miss. We will show how \name 
can improve bandwidth efficiency over these designs, in 
\cref{sec:eval}.

When a cacheline/page is inserted, 
%Data inserted to the DRAM cache may evict another piece of data in 
%order to make space. 
a replacement policy must select a victim cacheline/page.
Alloy Cache is direct mapped, and so only has one choice.
%In direct mapped caches,
%like Alloy Cache, there is no decision.
Conventional
set-associative caches (e.g., Unison Cache) use least-recently-used 
(LRU)~\cite{jevdjic2014} or frequency-based (FBR)~\cite{jiang2010} 
replacement.  These policies typically require additional 
metadata to track the relative age-of-access or access frequency for 
cachelines. Loading and updating the metadata incurs significant DRAM 
traffic. TDC implements a fully-associative DRAM cache but uses a FIFO 
replacement policy, which may hurt hit rate. Since Unison Cache and 
TDC do replacement at page granularity for each cache miss, they 
cannot support large pages efficiently. 

%A random policy avoids the need for additional metadata.

%\TODO{Figure out what Unison Cache does and mention it above.
%I looked and looked, and can't tell what it does.  Shameful
%on the part of the authors.  There are LRU bits, but they
%don't talk about updating them.  I suspect they cheated.}

\subsubsection{Software-Managed}

Software-implemented cache replacement algorithms (e.g., 
HMA~\cite{meswani2015}) can be fairly
% Software-managed caches rely on a software-implemented
%algorithm to determine what to place in the cache.
%These algorithms can be fairly 
sophisticated, and so may do
a better job than hardware mechanisms at predicting the
best data to hold in the cache.  However, they incur
significant execution time overhead, and therefore, are
generally invoked only periodically.  This makes them
much slower to adapt to changing application behavior.

\section{\name DRAM Cache Design}

%Since the main advantage of in-package DRAM over
%off-package DRAM is bandwidth rather than latency, 
\name aims to maximize bandwidth efficiency for both in- and 
off-package DRAM. To track DRAM contents, \name manages data mapping 
at page granularity using the page tables and TLBs like TDC and 
software-based designs. Different from previous designs, however, 
\name does not change a page's physical address when it is remapped.   
Extra bits are added to PTEs/TLBs to indicate whether the page is 
cached or not. This helps resolve the \textit{address consistency} 
problem (cf.  \cref{sec:remapping}). \name also uses a simpler and 
more efficient TLB coherence mechanism through software hardware 
co-design.

\subsection{\name Architecture} \label{sec:arch}

\begin{figure}[t!]
	\centering
    \includegraphics[width=0.9\columnwidth]{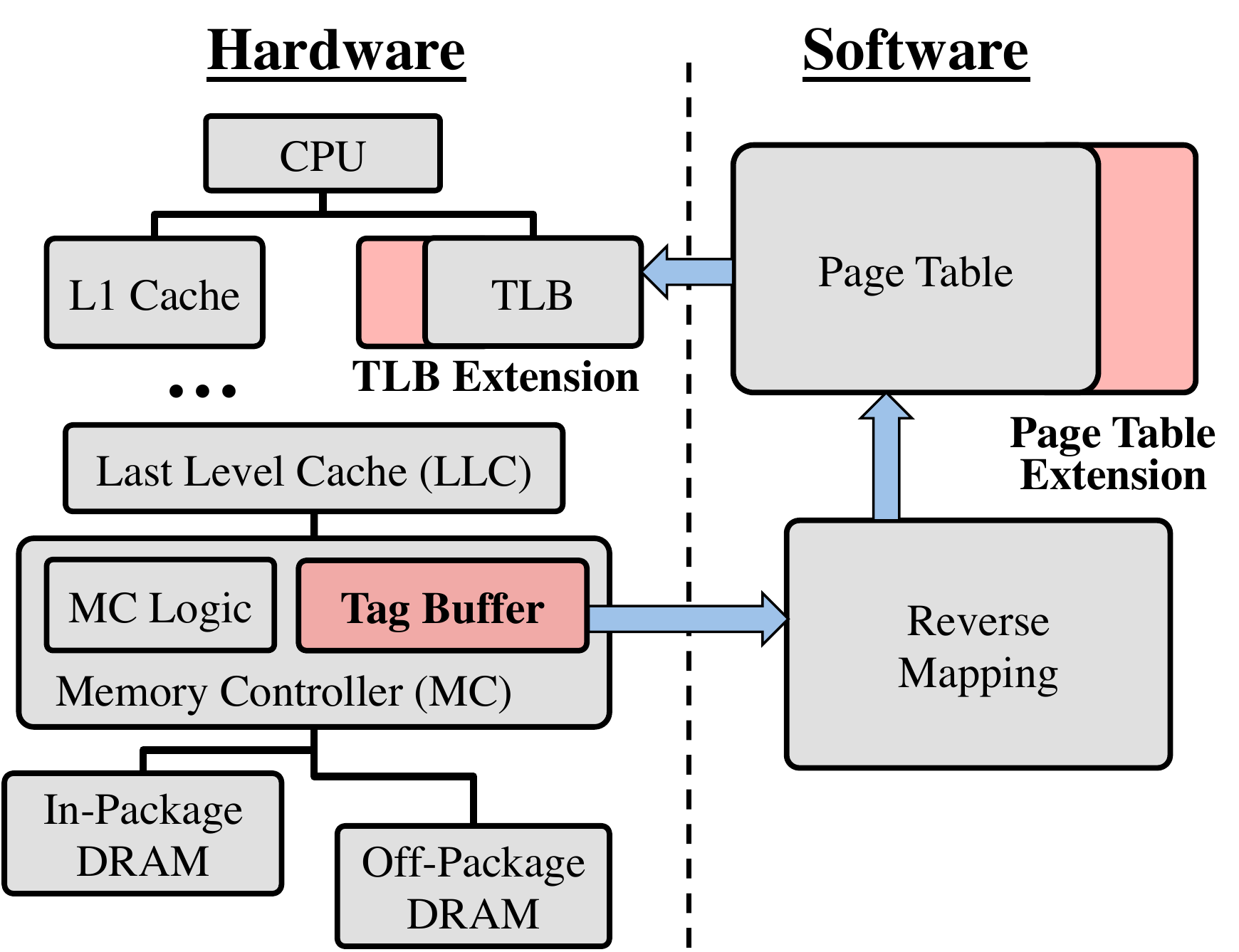}
	\caption{ Overall Architecture of \name. Changes to
	hardware/software components are highlighted in red.} 
	\vspace{-.2in}
    \label{fig:arch}
\end{figure}

%As discussed in \cref{sec:remapping}, previously proposed DRAM cache 
%schemes that rely on address remapping either fail to exploit 
%temporal locality at a fine grain, or incur complexity for TLB 
%coherence and address consistency (e.g., scrubbing caches on each 
%page remapping). 

\name implements a {\it lazy} TLB coherence protocol. 
%previous proposed software or hardware based TLB coherence mechanism 
%either fails to exploit locality at fine temporal granularity, or 
%incurs complexity and  undesirable side effects like the 
%\textit{address consistency} problem.  In \name, we combine the 
%benefits of software and hardware based designs and implement 
%use a new design which is a middleground between previous software 
%and hardware designs.  \name implements
%a \textit{lazy} TLB coherence mechanism
Information of recently remapped pages is managed in hardware and 
periodically made coherent in page tables and TLBs with software 
support. Unlike a software based solution, the cache replacement 
decision can be made in hardware and take effect instantly.  Unlike 
previous hardware based solutions, Banshee avoids the need for cache 
scrubbing. % which reduces the cost of TLB coherence.

Specifically, \name achieves this by adding a small hardware table in 
each memory controller. The table, called the \textit{Tag Buffer}, 
holds information on recently remapped pages that is not yet updated 
in the PTEs. When a page is inserted into or evicted from in-package 
DRAM, the tag buffer is updated but the corresponding PTEs and TLBs 
are not. Since all LLC misses to that page go through the memory 
controller, they will see the up-to-date mapping even if the request 
carries a stale mapping from a TLB.  Therefore, there is no need to 
update the TLBs eagerly. When the tag buffer eventually gets filled 
up, we push the latest mapping information to the PTEs and TLBs 
through a software interface.
Essentially, the tag buffer allows us to update the page table lazily
in batches, eliminating the need for fine-grained TLB coherence. 
%and therefore amortizes the cost.

%Another nice property of the tag buffer is that it decouples the tag 
%maintenance from the cache replacement policy. In principle, any 
%hardware managed replacement policy can be implemented in \name. We 
%will discuss our policy in \cref{sec:fbr}.

\cref{fig:arch} shows the architecture of \name. Changes made to both 
hardware and software (TLB/PTE extensions and the tag buffer) are 
highlighted in red. The in-package DRAM is a memory side cache and is 
not inclusive with respect to on-chip caches. 
%For clarity, only a single core and memory controller are shown.  In 
%practice, a system may contain multiple cores and memory controllers.  
We explain the components of the architecture in the rest of this 
section.

\subsection{PTE Extension} \label{sec:pte}

%\begin{figure}[t!]
%    \centering
%    \includegraphics[width=0.9\columnwidth]{figs/pte.pdf}
%    \caption{ Page Table Entry Format \TODO{show more
%    detail}}\label{fig:pte}
%\end{figure}

DRAM cache in \name is set-associative, each PTE is extended with 
mapping
information indicating whether (\textit{cached} bit) and where 
(\textit{way} bits) a page is cached. 
%We assume that the in-package DRAM is not a seperate part of the 
%physical address space, i.e., \name has inclusion between the in- and 
%off-package DRAM.
The \textit{cached} bit indicates whether a page is resident in DRAM 
cache, and if so, the \textit{way} bits indicate which way the page is 
cached in.

Every L1 miss carries the mapping information (i.e., cached bit and 
way bits) from the TLB through the memory hierarchy.
If the access is satisfied before it reaches a memory controller,
the cached bit and way bits are simply ignored. If 
the request misses the LLC and reaches a memory controller, it first 
looks up the tag buffer for the latest mapping. A tag buffer miss 
means the attached information is up-to-date.
For a tag buffer hit, the mapping carried by the request is ignored 
and the mapping info from the tag buffer is used.

% For an \textit{n}-way set associative cache, $\log{n}$ 
%bits are required to indicate the way number in a TLB or page table 
%entry.  
% CJH: The sentence above isn't wrong, but I think unnecessary.

% CJH: Moved this paragraph into section 2
%Some previous software-managed DRAM cache designs use different 
%physical address spaces for in-package and off-package DRAM.
%This NUMA-like architecture exposes more memory space to the OS.
%However, page remapping becomes 
%more expensive. If a page is inserted into or evicted 
%from an in-package DRAM, not only do we need to update the PTEs,
%but all the LLC cachelines belonging to the affected page 
%need to be updated/invalidated, since their physical addresses (or 
%cache addresses) have changed.

%\TODO{Explain how to add NUMA support for \name in this paragraph,
%or delete the whole paragraph.}

Unlike previous PTE/TLB-based designs which supports NUMA style DRAM 
cache (i.e., in- and off-package DRAMs have separate physical address 
space), \name assumes inclusion between in- and off-package DRAMs and 
access memory with a single address space. We make this design 
decision because the NUMA style caching will suffer from the 
\textit{address consistency} problem as discussed in 
\cref{sec:remapping}. Namely, whenever a page is remapped, all 
cachelines in on-chip caches belonging to the page need to be updated 
or invalidated for consistency. This incurs significant overhead when 
cache replacement is frequent.  In \name, however, remapping a page 
does not change its physical address, which avoids the address 
consistency issue.

%and reduces the overhead.  

%However, if such overhead is willing to be paid, \name can also 
%support NUMA with straightforward changes.  

%It is feasible to extend \name to support a NUMA-style DRAM cache, 
%where the cache is a separate part of the physical
%address space from off-package DRAM. However, as discussed in 
%\cref{sec:remapping}, this would incur extra overhead to maintain 
%address consistency.  Specifically, changing the mapping of a page 
%requires invalidating or updating all the lines in the LLC mapped to 
%that page.   In this paper, we stick with a single physical address 
%space for simplicity.

%\cref{fig:pts} shows the bits extension to a page table or TLB entry 
%in \name. Since most architectures have some unused bits in their page 
%table entries, we can use a few to represent the page mapping 
%information.   Specifically, a \textit{cached bit} is added to 
%indicate whether the page is cached in 3D-stacked DRAM or not. If 
%cached bit is set, then the \textit{way bits} indicate which way the 
%page is mapped to. For a core initiated request, the cached bit and 
%way bits are carried by request messages down the memory hierarchy.  
%They are used to decide whether the dram cache should be accessed and 
%if so, which way should be accessed.

Hardware prefetches from the L2 cache or lower present a complication.
These caches typically operate in physical address space, and thus 
cannot access TLBs for their mapping information. 
%thus do not have a TLB. As a result, the mapping information for the 
%prefetched address cannot be looked up.
In most systems, however, prefetches of this sort stop at a page
boundary, since the data beyond that boundary in physical address space
is likely unrelated to the previous page. %; we assume such a design.
Further, these prefetches are always triggered
(directly or indirectly) by demand or prefetch requests coming from
the core or L1.
Thus, we can copy the mapping information from a triggering access to
all prefetches it triggers.

\subsection{Tag Buffer} \label{sec:tagbuffer}

\begin{figure}[t!]
	\centering
    \includegraphics[width=0.95\columnwidth]{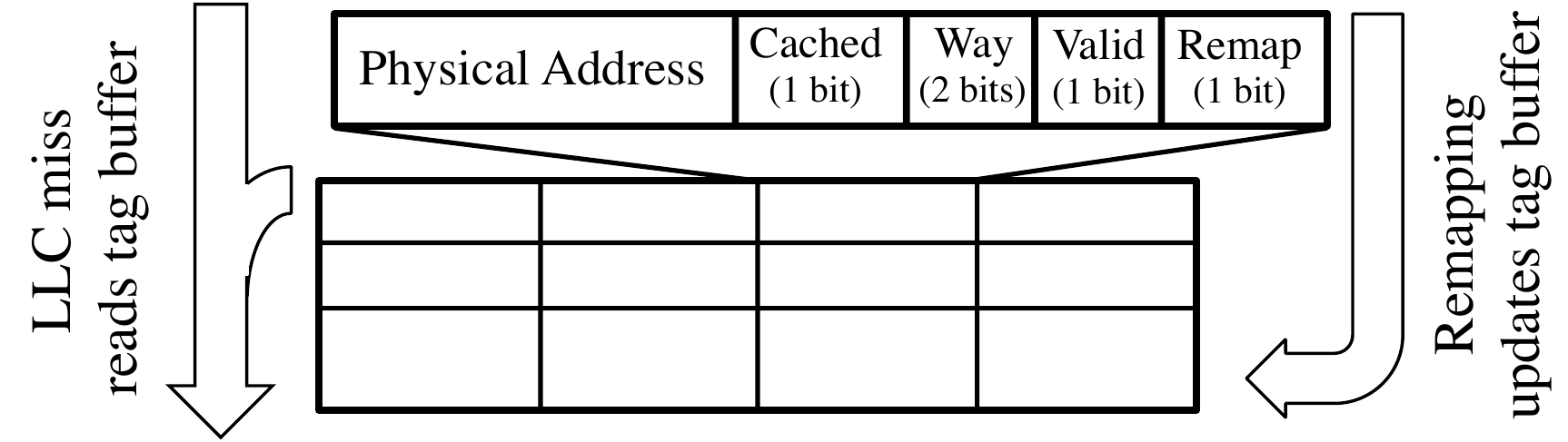}
    \vspace{-.1in}
    \caption{ Tag buffer organization. }
	\vspace{-.2in}
    \label{fig:tagbuffer}
\end{figure}

A tag buffer resides in each memory controller and holds the mapping 
information of recently remapped pages belonging to
that memory controller.  \cref{fig:tagbuffer} shows the 
architecture of a tag buffer. It is organized as a set associative 
cache with the physical address as the tag. 
%The mapping information stored in the tag buffer (cached bits and way 
%bits) has the same format as that in the page table or TLB. Each 
%request arriving at the
%memory controller looks in the tag buffer for the latest mapping, and 
%each cache replacement updates the tag buffer.
The \textit{valid} bit indicates whether the entry contains a valid 
mapping. For a valid entry, the \textit{cached} bit and \textit{way} 
bits indicate whether and where the page exists in DRAM cache. The 
\textit{remap} bit is 1 if the mapping is not yet reflected in the 
page tables. 

Most requests arriving at a memory controller carry mapping information,
except for LLC dirty evictions. 
%since the mapping bits are not stored in the LLC.
If the mapping of the evicted cacheline is not in the tag buffer, then 
the memory controller needs to probe the tags stored in the DRAM cache 
(cf.  \cref{sec:layout}) to 
%--- the memory controller reads the tags for the evicted line's set, 
%and compares them with the line's address to 
determine if this is a hit or miss.
These probing operations consume DRAM cache bandwidth. 

To reduce such tag probes for dirty eviction, we use otherwise empty 
entries in the tag buffer to hold mappings for pages cached in the 
LLC. On LLC misses that also miss in the tag buffer, we
allocate an entry in the tag buffer for the page. While the
\textit{valid} bit is set to 1, indicating a useful mapping,
the \textit{remap} bit is set to 0, indicating the entry stores
the same mapping as in the PTEs.
Such entries can be replaced in the tag buffer without affecting 
correctness. %, unlike entries with \textit{remap} set.
We use an LRU replacement policy among entries with \textit{remap} 
unset, which can be implemented by running the normal LRU algorithm with 
the remap bits as a mask.

%CJH: I wouldn't bother bringing the below up
%\red{TODO. (optional) another way to get rid of tag probing for LLC 
%dirty eviction is to maintain the mapping info for cachelines in the 
%LLC.  This means for each page remapping, all the affected lines in 
%the LLC needs to be updated. Since cachelines belonging to the same 
%page may be mapped to different LLC slices, this solution incurs too 
%much on chip traffic for each remapping.  }

\subsection{Page Table and TLB Coherence} \label{sec:coherence}

As the tag buffer fills, the mapping information stored in it needs to 
be migrated to the page table, to make space for future cache 
replacements. Since the tag buffer only contains the physical address 
of a page but page tables are indexed using virtual addresses, we need 
a mechanism to identify all the PTEs corresponding to a physical 
address.  

TDC has proposed a hardware inverted page table
to map a page's physical address to its PTE~\cite{lee2015}. This 
solution, however, cannot handle the \textit{page aliasing} problem 
where multiple virtual addresses are mapped to the same physical 
address. 
%The problem is tricky since when a page is remapped, it is even hard 
%to know whether aliasing exists or not, since such information is not 
%included in the PTE.  
To figure out whether aliasing exists, some internal structure in an 
OS (i.e., page descriptors) has to be accessed which incurs 
significant extra overhead.  

\begin{figure*}[t!]
    \centering
    \includegraphics[width=0.85\textwidth]{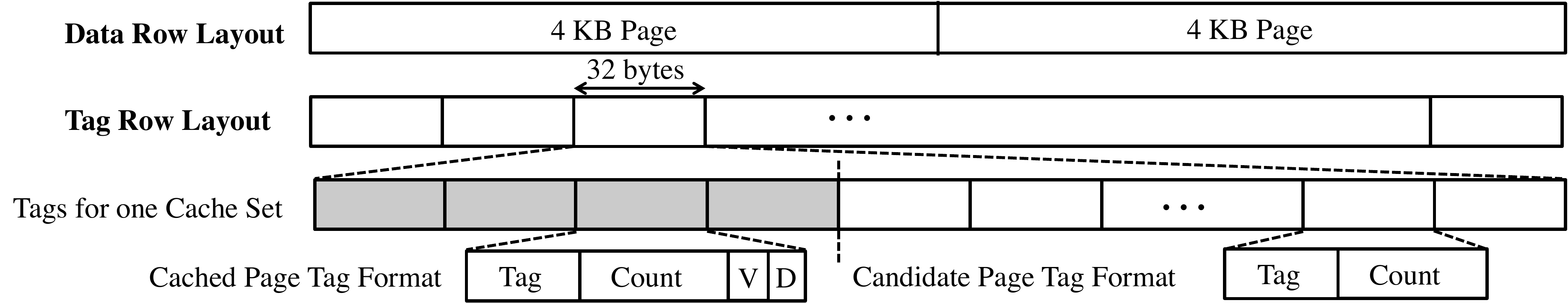}
    \caption{ 4-way associative DRAM cache layout (not drawn to 
    scale).}
    \vspace{-.2in}
    \label{fig:layout}
\end{figure*}

We observe, however, that a modern OS already has a \textit{reverse 
mapping} mechanism to quickly identify the associated PTEs for a 
physical page, regardless of any aliasing.
This functionality is necessary to 
implement page replacement between main memory and secondary storage 
(e.g., Disk or SSD) since reclaiming a main memory page frame requires 
accessing all the PTEs mapped to it.  Reverse mapping can be 
implemented through an inverted page table (e.g., Ultra SPARC and 
Power PC~\cite{stallings1998}) or a special reverse mapping mechanism 
(e.g., Linux~\cite{bovet2005}). In \name, the PTE coherence scheme is 
implemented using reverse mapping. 

When a tag buffer fills up to a pre-determined threshold, it sends
an interrupt to one or 
more cores. The core(s) receiving the interrupt will 
execute a software routine to update recently remapped pages'
PTEs.  Specifically, all entries are read from the tag buffers in all 
memory controllers (which are memory mapped). For each tag buffer 
entry, the physical address is used to identify the corresponding PTEs 
through the reverse mapping mechanism. Then, the \textit{cached} bit 
and \textit{way} bits are updated for each PTE. During this process, 
the tag buffers can be locked so that no DRAM cache replacement 
happens.  But the DRAMs can still be accessed and no programs need to 
stopped.
%programs in the system do not need to stop.  

After all tag buffer entries have been applied to the page table, the 
software routine issues a system wide TLB shootdown to enforce TLB 
coherence. After this, a message is sent to all tag buffers to clear 
the \textit{remap} bits for all entries.  Note that the mapping 
information can stay in the tag buffer to help reduce tag probing for 
dirty evictions (cf.  \cref{sec:tagbuffer}).

Depending on a system's software and hardware, the
mechanism discussed above may take many cycles. However, since 
this cost only needs to be paid once a tag buffer is almost 
full, the cost of updating PTEs is amortized. Furthermore, as we will see in 
\cref{sec:fbr}, remapping pages too often leads to poor performance
due to high replacement traffic. Thus, our design tries
to limit the frequency of page remapping, further reducing the cost
of PTE updates.

\section{Bandwidth-Efficient Cache Replacement} \label{sec:fbr}

As discussed in \cref{sec:background-replacement}, the cache 
replacement policy can significantly affect traffic in DRAMs. This is 
especially true for page granularity DRAM cache designs due to the 
over fetching problem.
%where a single access to a cacheline may cause a whole page of data 
%be moved between DRAMs.  The impact of each replacement decision 
%increases with the granularity of the data held in the cache; since 
%\name uses page granularity, it benefits greatly from an improved 
%policy.
In this section, we propose a \textit{frequency based} replacement 
(FBR) policy with sampling to achieve a good hit rate while minimizing 
DRAM traffic.

We first discuss the physical layout of the data and metadata
in the DRAM cache in
\cref{sec:layout}.  We then describe \name's
cache replacement algorithm in \cref{sec:repl}.

\subsection{DRAM Cache Layout} % and Counter Organization}
\label{sec:layout}

Many previously proposed tag-based DRAM cache schemes store 
the tags and data in the same DRAM row to exploit row buffer 
locality, since they always access tags along with data.
Such an organization can be efficient for a cacheline
granularity DRAM cache.  For a page granularity DRAM cache, however, 
pages and tags do not align well within a DRAM row 
buffer~\cite{jevdjic2014}, which incurs extra design complexity and 
inefficiency.

In \name, the tags are rarely accessed --- only for cache replacement 
and LLC dirty evictions that miss in the tag buffer. Therefore, tags 
and data are stored separately for better alignment.  
%there is little downside to storing tags and data separately.
\rev{\cref{fig:layout} shows the layout of a data row and a tag row in 
a DRAM cache with row buffer size of 8 KB and page size of 4 KB.}  The 
tags and other metadata of each DRAM cache set take 32 bytes in a tag 
row.  For a 4-way associative DRAM cache, each set contains 16~KB of 
data and 32~bytes of metadata, so the metadata overhead is only 0.2\%.  

%\subsubsection{Frequency Counter Storage}

\name tracks each page's access frequency with a counter, stored
in the metadata. 
%Since we will use access counts to make replacement decisions, 
We store counters not only for the pages in the DRAM cache, but also 
for some pages \textit{not} in cache, which are candidates to bring
into the cache.
Intuitively, we want to cache pages that are most frequently accessed, 
and track
pages that are less frequently accessed as candidates.

%\cref{fig:layout} shows the layout of set metadata assuming a 4-way 
%associative cache.
%The first part of each set's metadata contains information on pages 
%in the cache: tags, frequency counters, valid, and dirty bits.  The 
%second part contains information on candidates for caching that map 
%to the set. For candidate pages, \name stores only tags and frequency 
%counters.
%for caching by maintaining their frequency information.   

\subsection{Bandwidth Aware Replacement Policy} \label{sec:repl}

A frequency-based replacement policy incurs DRAM cache traffic through 
reading and updating the frequency counters and through replacing 
data.
In \cref{sec:sample}, we introduce a 
sampling-based counter maintenance scheme to reduce the counter 
traffic. In \cref{sec:repl-alg}, we discuss the bandwidth aware 
replacement algorithm that attempts to minimize replacement
traffic while maximizing hit rate.   

\subsubsection{Sampling-Based Counter Updates} \label{sec:sample}

In a standard frequency-based replacement policy~\cite{lee2001, 
robinson1990},  each access increments the data's frequency counter.
We observe, however, that 
incrementing the counter for each access is not necessary.  Instead, 
an access in \name only updates a page's frequency counter with a 
certain \textit{sample rate}.
%Specifically, we generate a random number between 0 and 1
%on each access, and only update the counter if the number
%is less than the sample rate.  
For a \textit{sample rate} of 10\%, for example, the frequency 
counters are accessed/updated only once for every 10 DRAM accesses.  
This will
reduce counter traffic by 10$\times$.
Furthermore, since sampling slows the incrementing of the counters,
we can use fewer bits to represent each counter. 
%without losing information.

It may seem that updating counters based on sampling leads 
to inaccurate detection of ``hot'' pages. However, the vast majority
of applications exhibit significant spatial locality. When a cacheline 
misses in the DRAM cache, other cachelines belonging to the same page 
are likely to be accessed soon as well. Each of these accesses to the 
same page has a chance to update the counter. In fact, without 
sampling, we find that counters quickly reach large values but only 
the high order bits are used for replacement decision. Sampling 
effectively discards the low-order bits of each counter,
which have little useful information anyway.

We further observe that when the DRAM cache works well, i.e., it has 
low miss rate, replacement should be rare and the counters need not be 
frequently updated.  Therefore, \name uses an
adaptive sample rate which is the product of the cache miss rate
and a constant rate (\textit{sampling coefficient}).

%\TODO{When do we allocate counters for candidates?}

%\TODO{When do we decrease the counters?  I believe we can say
%something like the following, although more detail needs to be filled 
%in. To keep counters from saturating, and to ensure counters 
%represent recent application behavior, \name periodically reduces all 
%counters in a set by half, via a shift operation.  This is done only 
%when we are ready to do a counter update on the set anyway, to 
%eliminate any extra traffic.
%}

%In \name, the sample rate is tuned by the cache 
%miss rate. Specifically, the actual sample rate is the product of a 
%static sample rate (fixed in hardware) and the recent miss rate of the 
%DRAM cache.
%
%\[ dynamicSampleRate = staticSampleRate \times missRate \]

\subsubsection{Replacement Algorithm} \label{sec:repl-alg}

DRAM cache replacement can be expensive, in terms of traffic, 
especially for page granularity designs. 
%and this is exacerbated when the cache is managed at page 
%granularity.  
For each replacement, the memory controller transfers a whole page 
(assuming no footprint cache) from off-package DRAM to in-package 
DRAM.  Even worse, if the evicted
page is dirty, the memory controller must transfer it from in-package 
DRAM to off-package DRAM, doubling the traffic for the replacement.
For cases where a page sees only a few accesses before being replaced,
we may actually see higher off-package DRAM traffic (and worse performance)
than if the DRAM cache was not present.

Frequency-based replacement does not inherently preclude this problem.  
In a pathological case for FBR, we may keep replacing the least 
frequently accessed page in the cache with a candidate whose counter 
has just exceeded it. When pages have similar counter values, a large 
number of such replacements can be triggered, thrashing the cache and 
wasting bandwidth.  
%cache full of lines with very small counter values, and
%a replacement on most accesses.

\name solves this problem by only
%using a simple solution to limit thrashing.
%Specifically, it only
replacing a page when the candidate's counter is greater than the 
victim's counter \textit{by a certain threshold}.
This ensures that a page just evicted from the DRAM cache must be 
accessed for at least $\frac{2 \cdot threshold}{{sampling}~{rate}}$ 
times before it can enter the cache again,
thus preventing a page from entering and leaving frequently.
Note that reducing the frequency of replacement also increases the 
time between tag buffer overflows, indirectly reducing the overhead of
updating PTEs.

\setlength{\textfloatsep}{10pt}
\begin{algorithm}[t]
    \footnotesize
    \textbf{Input} : tag \\ %, cache\_hit}
    \codeComment{rand(): random number between 0 and 1.0} \\
    \If{rand() < recent\_miss\_rate $\times$ sampling\_coeff}
    { meta = dram\_cache.loadMetadata(tag) \\
      \eIf{tag in meta}{
        meta[tag].count ++ \\
        \If{tag in meta.candidates \textbf{and} meta[tag].count 
        > meta.cached.minCount() + threshold}{
			replace the cached page having the minimal counter with 
	  the accessed page}
		\If{meta[tag].count == max\_count} {
			\codeComment{Counter overflow, divide by 2} \\
            \ForAll{t in meta.tags} {
                meta[t].count /= 2
            }
        }
        dram\_cache.storeTag(tag, metadata) \\
      } {
        victim = random page in meta.candidates \\
        \If{rand() < 1 / victim.count}{
          victim.tag = tag \\
          victim.count = 1 \\
          dram\_cache.storeTag(tag, metadata) \\
        }
      }
    }
    \caption{Cache Replacement Algorithm}
    \label{alg:replacement}
\end{algorithm}

The complete cache replacement algorithm of \name is shown in 
Algorithm~\ref{alg:replacement}. For each request from the LLC, a 
random number is generated to determine whether the current access 
should be sampled. If it is not sampled, which is the common case, 
then the access is made to the proper DRAM (in- or off-package) 
directly.  No metadata is accessed and no replacement happens. 
%This is the key to why \name is able to achieve good bandwidth 
%efficiency.

If the current access is sampled, then the metadata for the 
corresponding set is loaded from the DRAM cache to the memory 
controller. If the currently accessed page exists in the metadata, its 
counter is incremented.  Furthermore, if the current page is in the 
candidate part and its counter is
greater than a cached page's counter by a threshold, then cache 
replacement should happen.  By default, the threshold is the 
product of the number of cachelines in a page and the sampling 
coefficient divided by two (\textit{threshold} $=$ \textit{page\_size} 
$\times$ \textit{sampling\_coeff} / 2).  Intuitively, this means 
replacement can happen only if the benefit of swapping the pages 
outweighs the cost of the replacement operation.  If a counter 
saturates after being incremented, all counters in the metadata will 
be reduced by half using a shift operation in hardware. 

If the current page does not exist in the metadata, then a random page 
in the candidate part is selected as the victim. The current page can 
overtake the victim entry with a certain probability, which decreases 
as the victim's counter gets larger.  This way, it is less likely that 
a hot candidate page is evicted.  

\subsection{Supporting Large Pages}

Large pages have been widely used to reduce TLB misses and therefore 
should be supported in DRAM caches.  In \name, we manage large pages 
using PTEs and TLBs as with regular pages.  We assume huge pages 
(1~GB) are managed purely in software and discuss the hardware support 
for large pages (2~MB) here.  

In \name, the DRAM cache can be partitioned to two portions for normal 
and large pages respectively. Partitioning can happen at context 
switch time by the OS which knows how many large pages each process is
using. Partitioning can also be done dynamically using runtime 
statistics based on access counts and hit rates for different page 
sizes. Since most of our applications either make very heavy use of 
large pages, or very light usage, partitioning could give either most 
or almost none of the cache, respectively, for large pages. We leave a 
thorough exploration of these partitioning policies for future work.  

We force each page (regular or large) to map to a single MC (memory 
controller) to simplify the management of frequency counters and cache 
replacement. A memory request learns the size of the page being 
accessed from the TLB, and uses this information to determine which MC 
it should access. In order to figure out the MC mapping for LLC dirty 
evictions, a bit is appended to each on-chip
cacheline to indicate its page size. When the OS reconfigures large 
pages, which happens very rarely~\cite{huge-page}, all lines within 
the affected pages should be flushed from the LLC and in-package 
DRAMs. 

In terms of the data and tag layout, a large page mapped to a
particular way will span multiple cache sets taking the corresponding 
way in each set. One difference between regular and large pages is the 
cache replacement policy. Due to the higher cost of replacing a large 
page, the frequency counters need a greater threshold for replacement.  
We also reduce the sample rate of updating frequency counters to 
prevent counter overflow. Note that large pages do not work well for 
page-granularity schemes that replace on each DRAM cache miss.  TDC, 
for example, disables caching of large pages. 

\section{Evaluation} \label{sec:eval}

We now evaluate the performance of \name and compare it to
other DRAM cache designs. \cref{sec:meth} discusses the methodology of 
the experiments. 
%\cref{sec:eval-tag} shows how \name reduces the tag overhead.  
\cref{sec:eval-main} and \cref{sec:eval-traffic} show the performance 
and DRAM traffic comparison of different DRAM cache designs. Finally, 
\cref{sec:eval-sensitivity} presents sensitivity studies.

\subsection{Methodology} \label{sec:meth}

\begin{table}[!t]
    \caption{ System Configuration. }
    \vspace{-0.2in}
    \begin{center}
    { \footnotesize
        \begin{tabular}{|l|l|}
            \hline
			\multicolumn{2}{|c|}{System Configuration} \\
            \hline
            Frequency					& 2.7~GHz \\
            Number of Cores				& N = 16 \\
            Core Model                  & 4-Issue, Out-of-Order \\
			\hline
            \hline
			\multicolumn{2}{|c|}{Memory Subsystem} \\			
            \hline
			Cacheline Size        		& 64~bytes \\
            L1 I Cache					& 32~KB, 4-way \\
            L1 D Cache                	& 32~KB, 8-way \\
            L2 Cache                    & 128~KB, 8-way \\
            Shared L3 Cache             & 8~MB, 16-way \\
            %Memory Controllers (MC)     & 1 \\
            \hline
            \hline

            \multicolumn{2}{|c|}{Off-Package DRAM} \\
            \hline
            Channel                     & 1 channel \\
            Bus Frequency               & 667 MHz (DDR 1333 MHz)\\
            Bus Width                   & 128 bits per channel \\
            tCAS-tRCD-tRP-tRAS          & 10-10-10-24 \\
            \hline
            \hline
            \multicolumn{2}{|c|}{In-Package DRAM} \\
            \hline
            Capacity                    & 1~GB \\
            Channel                     & 4 channels \\
            Bus Frequency               & 667 MHz (DDR 1333 MHz)\\
            Bus Width                   & 128 bits per channel \\
            tCAS-tRCD-tRP-tRAS          & 10-10-10-24 \\
            \hline
		\end{tabular}
    }
    \end{center}
    \label{tab:system}
    \vspace{-0.2in}
\end{table}

\begin{table}
    \caption{ \name Configuration. }
    \vspace{-0.2in}
    \begin{center}
    { \footnotesize
        \begin{tabular}{|l|l|}
            \hline
            \multicolumn{2}{|c|}{DRAM Cache and Tag} \\
            \hline
            Ways                        & 4 \\
            Page Size                   & 4096~KB \\
            Tag Buffer                  & 1 buffer per MC \\
                                        & 8-way, 1024 entries \\
                                        & Flushed when 70\% full \\
            Tag Buffer Flush Overhead   & 20~us  \\
            TLB Shoot Down Overhead     & Initiator 4~us, slave 1~us 
            \\
            \hline
            \multicolumn{2}{|c|}{Cache Replacement Policy} \\
            \hline
            Cache Set Metadata          & 4 cached pages \\
                                        & 5 candidate pages \\
            Frequency Counter           & 5 bits \\
            Sampling Coefficient        & 10\% \\
            \hline
        \end{tabular}
    }
    \end{center}
    \label{tab:hybrid}
    \vspace{-0.1in}
\end{table}

We use ZSim~\cite{sanchez2013} to simulate a multi-core processor 
whose configuration is shown in \cref{tab:system}. The chip has
one channel of off-package DRAM and four channels of in-package DRAM.
We assume all the channels are the same to model behavior of 
in-package DRAM~\cite{hbm2014, knl-micro}. The maximal bandwidth that 
this configuration offers is 21~GB/s for off-package DRAM and 85~GB/s
for in-package DRAM. In comparison,
Intel's Knights Landing~\cite{knl2015} has roughly 4$\times$
the bandwidth and number of cores (72 cores, 90 GB/s off-package DRAM 
and 300+ GB/s in-package DRAM bandwidth), so we use the same bandwidth 
per core.

The default parameters of \name are shown in \cref{tab:hybrid}.  
\rev{Each PTE and TLB entry is extended with 3 bits for the mapping 
information.  This is a small storage overhead (4\%) for TLBs and zero 
storage overhead for PTEs since we are using otherwise unused bits.  
Each request in the memory hierarchy carries the 3 mapping bits.  Each 
memory controller has an 8-way set associative tag buffer with 1024 
entries, requiring only 5~KB storage per memory controller.} Hardware 
triggers a ``tag buffer full'' interrupt
when the buffer is 
70\% full. We assume the interrupt handler runs on a single randomly
chosen core and takes 20 microseconds.
For TLB shootdown, the initiating core pays an overhead of 4 
microseconds and every other core pays 1 microsecond 
overhead~\cite{villavieja2011}. 

The frequency counters are 5 bits long.
The 32-byte per set metadata holds
information for 4 cached pages and 5 candidate pages\footnote{With a
48-bit address space and the DRAM cache parameters, the tag size is 48 
- 16 ($2^{16}$ sets) - 12 (page offset) = 20 bits. Each cached page
has metadata of 20 + 5 + 1 + 1 = 27 bits and each candidate page has 
25 bits of metadata (\cref{fig:layout}).}.  The default sampling 
coefficient is 10\%  -- the actual sample rate is this multiplied by 
the recent DRAM cache miss rate.  
%In some experiments, we also show results with sampling coefficients 
%of 1 and 0.01, labeled appropriately.

\subsubsection{Baselines}

We compare \name to the following baselines.

\textbf{No Cache}: The system only contains off-chip DRAM.

\textbf{Cache Only}: The system only contains in-package DRAM with
infinite capacity. 
%This sets a loose upper bound on performance.

\textbf{Alloy Cache}~\cite{qureshi2012}: A state-of-the-art 
cacheline-granularity design, described in \cref{sec:back}.
We also include the \textit{bandwidth 
efficient cache fills} and the \textit{bandwidth efficient writeback 
probe} optimizations from BEAR~\cite{chou2015} to improve
bandwidth efficiency.  This includes a stochastic replacement
mechanism that only does replacement with 10\% probability.
In some experiments, we show results from always replacing
(\texttt{Alloy 1}), and replacing 10\% of the time (\texttt{Alloy 
0.1}).

\textbf{Unison Cache}~\cite{jevdjic2014}: A state-of-the-art 
page-granularity design, described in \cref{sec:back}.
We model an LRU replacement policy.  We assume perfect
way prediction and footprint prediction. For footprint prediction,
we first profile each workload to collect the average number of blocks 
touched per page fill; the actual experiments charge this amount of 
replacement traffic. The footprint is managed at 4-line granularity.  
We assume the predictors incur no overhead.

\textbf{Tagless DRAM Cache (TDC)}~\cite{lee2015}: A state-of-the-art 
page-granularity design described in \cref{sec:back}. We modeled an
idealized TDC configuration. Specifically, we assume a zero-overhead 
TLB coherence mechanism and ignore all the side effects of the 
mechanism (i.e., address consistency, page aliasing). We also 
implement a perfect footprint cache for TDC like we do for Unison 
Cache.

\begin{table}
    \caption{ Mixed SPEC workloads. }
    \vspace{-0.25in}
    \begin{center}
	%{ \scriptsize
	{ \footnotesize
        \begin{tabular}{|c|l|}
        \hline
        Name & Workloads \\
        \hline
        \hline
        Mix1 & libq-mcf-soplex-milc-bwaves-lbm-omnetpp-gcc $\times$ 2 
        \\
        \hline
        Mix2 & libq-mcf-soplex-milc-lbm-omnetpp-gems-bzip2 $\times$ 2
        \\
        \hline
        Mix3 & mcf-soplex-milc-bwaves-gcc-lbm-leslie-cactus $\times$ 2 
        \\
        \hline
        \end{tabular}
    }
    \end{center}
    \label{tab:mixture}
    \vspace{-0.2in}
\end{table}

\subsubsection{Benchmarks} \label{sec:eval-bench}

We use SPEC CPU2006~\cite{henning2006} and graph
analytics benchmarks~\cite{yu2015}. Each experiment is simulated for 
100 billion instructions or to completion, whichever happens first. By 
default, all benchmarks use small pages only.  

We selected a subset of SPEC benchmarks that have large memory 
footprint and consider both homogeneous and heterogeneous workloads.  
For homogeneous workloads, each core in the simulated system executes 
one
instance of a benchmark and all the instances run in parallel.
\rev{Heterogeneous workloads model the multi-programming environment 
where the cores run a mixture of benchmarks.
We use three randomly selected mixtures, shown in \cref{tab:mixture}}.

To represent throughput computing workloads, the target applications 
for the first
systems employing in-package DRAM, \rev{we include multi-threaded 
graph analytics workloads}.  We use all graph workloads 
from~\cite{yu2015}, which
span the spectrum of memory and compute intensity.

\begin{figure*}[t!]
    \centering
    \includegraphics[width=.9\textwidth]{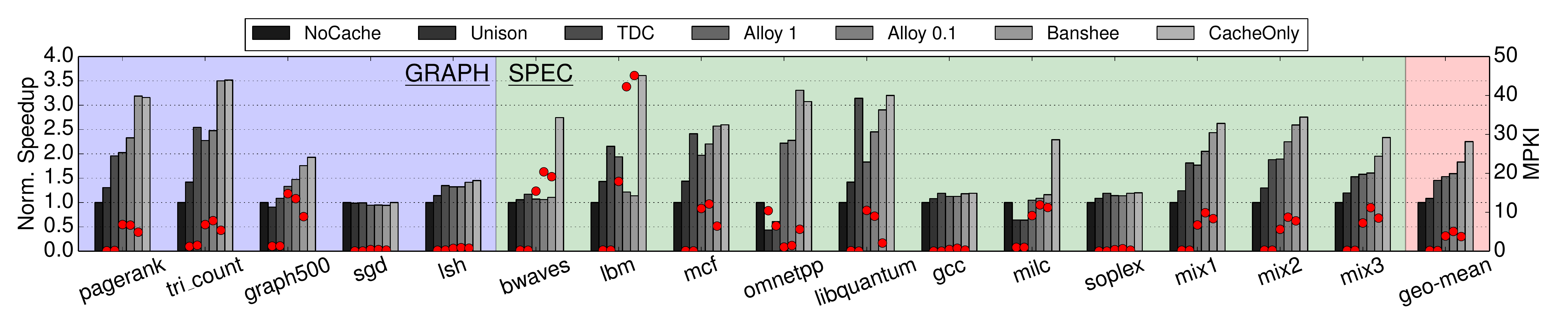}
    \vspace{-.15in}
	\caption{ Speedup normalized to NoCache. }
    \vspace{-.17in}
    \label{fig:speedup}
\end{figure*}

\begin{figure*}[t!]
    \centering
    \includegraphics[width=.9\textwidth]{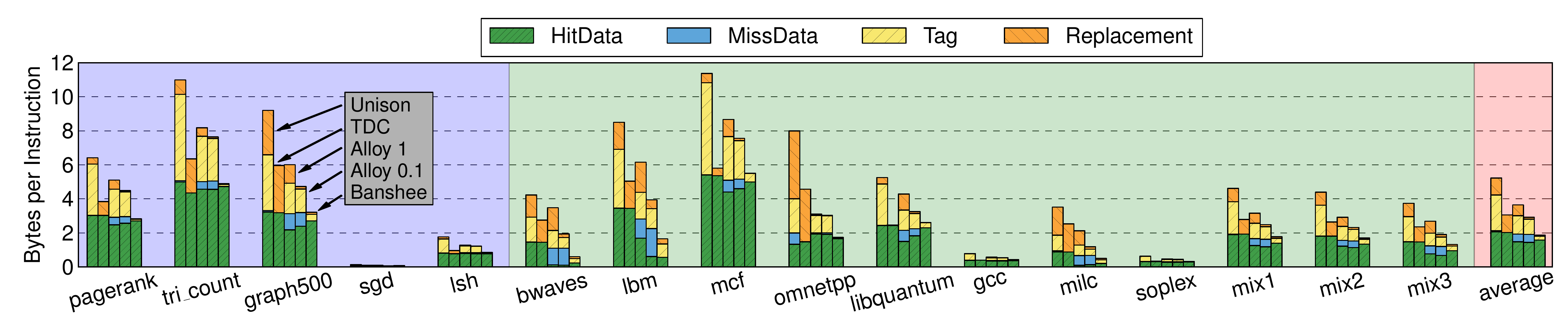}
    \vspace{-.1in}
	\caption{ In-package DRAM traffic.  }
    \vspace{-.17in}
    \label{fig:mc-traffic}
\end{figure*}

\begin{figure*}[t!]
    \centering
    \includegraphics[width=.9\textwidth]{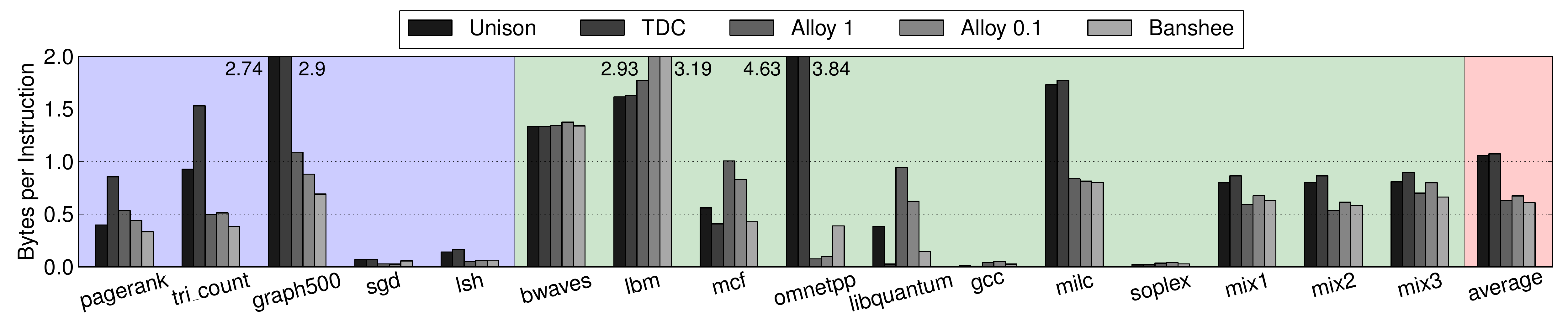}
	\vspace{-.1in}
    \caption{ Off-package DRAM traffic.  }
    \vspace{-.22in}
    \label{fig:ext-traffic}
\end{figure*}

Many benchmarks that we evaluated have very high memory bandwidth 
requirement. With the \texttt{CacheOnly} configuration, for example,
10 out of the 16 benchmarks have an average DRAM bandwidth consumption 
over 50 GB/s (bursts may exceed this). This bandwidth requirement 
exerts enough pressure to in-package DRAM (with maximum bandwidth of 
85~GB/s) so that bandwidth changes can significantly affect 
performance.  Our memory intensive benchmarks experience 2--4$\times$ 
higher memory access latency compared to compute intensive benchmarks 
due to the bandwidth bottleneck.

\subsection{Performance} \label{sec:eval-main}

%In this section, we present end-to-end performance evaluation of \name 
%with respect to baseline algorithms. 

%\TODO{Is the average geomean?  If so, we should say, ``The average
%bars indicate geometric mean across all workloads.''}

% The \texttt{CacheOnly} configuration assumes 
%infinite-bandwidth in-package DRAMs and is therefore the performance 
%upperbounds.   

\cref{fig:speedup} shows the speedup of different cache designs
normalized to \texttt{NoCache}. The average bars indicate geometric 
mean across all workloads. On average, \name provides a 68.9\% speedup 
over Unison Cache, 26.1\% over TDC and 15.0\%  over Alloy Cache.  The 
higher bandwidth efficiency is the major contributor to performance 
improvement. Compared to Unison Cache and Alloy Cache, \name can also 
reduce the cache miss latency since the DRAM cache need not be probed 
to check presence. 

Unison Cache and TDC have worse performance than other designs on some 
benchmarks (e.g., \texttt{omnetpp} and \texttt{milc}) due to the lack 
of spatial locality.  
%The performance can be even worse than the \texttt{NoCache} baseline.  
%This is due to the lack of spatial locality in these benchmarks.  
As a result, they spend a lot of DRAM traffic for cache replacement.   
Having a footprint predictor does not completely solve the problem 
since the footprint cannot be managed at cacheline granularity due to 
the storage overhead (we modeled 4-line granularity).  
%cache capacity
%compared to a block granularity one, leading to a higher miss rate,  
%even with the presence of a footprint predictor.
\name is also at page granularity, but its bandwidth-aware replacement 
policy offsets this inefficiency for these benchmarks.

On \texttt{lbm}, however, both \name and \texttt{Alloy 0.1} give
worse performance than other baselines. \texttt{lbm} has very good 
spatial locality on each page, but a page is only accessed a small 
number of times before it gets evicted.
\texttt{Alloy 1}, Unison Cache and TDC have good performance on 
\texttt{lbm} since they do replacement for every DRAM cache miss, 
therefore exploiting more locality.  \name and \texttt{Alloy 0.1}, in 
contrast, cannot leverage all the locality due to their selective data 
caching.
One solution is to dynamically switch between 
different replacement policies based on a program's access pattern.  
For example, some pre-determined sets in the cache may use 
different replacement policies and hardware selects the policy for
the rest of the cache through set dueling~\cite{chou2015, 
qureshi2007}.  We leave exploration of this for future work.

%To help explain the performance results,
The red dots in \cref{fig:speedup} shows the \textit{Miss Per Kilo 
Instruction} (MPKI) for each DRAM cache scheme on different 
benchmarks. In general, Alloy Cache and \name achieve
similar miss rates, while Unison Cache and TDC have a very low miss 
rate since we assume perfect footprint prediction for them.

%For Alloy Cache, while stochastic replacement (\texttt{Alloy 0.1}) 
%reduces DRAM cache traffic, it gives a higher miss rate, leading to 
%slightly worse average performance.

For some benchmarks (e.g., pagerank, omnetpp), \name performs even 
better than \texttt{CacheOnly}. This is because \texttt{CacheOnly} has 
no external DRAM. So the total available DRAM bandwidth is less than 
\name which has both in- and off-package DRAM. We will have more 
discussion of balancing DRAM bandwidth in \cref{sec:balance}.   

%\begin{figure}[t!]
%    \centering
%    \includegraphics[width=\columnwidth]{figs/mc_avg.pdf}
%    \caption{ Normalized DRAM cache traffic breakdown. }
%    \label{fig:mc}
%\end{figure}
%
%\begin{figure}[t!]
%    \centering
%    \includegraphics[width=\columnwidth]{figs/ext_avg.pdf}
%    \caption{ Normalized off-package DRAM traffic breakdown. }
%    \label{fig:ext}
%\end{figure}

%\TODO{Why does the y-axis say Load Miss Rate?  Does this ignore
%stores?}

\subsection{DRAM Traffic} \label{sec:eval-traffic}

\cref{fig:mc-traffic} and \cref{fig:ext-traffic} show the in- and 
off-package DRAM traffic respectively.  \rev{Traffic is measured
in \textit{bytes per instruction} to convey memory intensity of a 
workload, in addition to comparative behavior of the cache designs.}

In \cref{fig:mc-traffic}, the \texttt{HitData} is the data transfer 
for DRAM cache hits, which is the only useful data transfer; 
everything else is overhead. For Alloy and Unison Cache, 
\texttt{MissData} is the speculative data loading for cache miss and 
\texttt{Tag} is the traffic for tag accesses. \texttt{Tag} also 
represents the frequency counter accesses and tag probes for LLC dirty 
evictions in \mbox{\name}. 
%Both have small traffic and performance overhead.  
\texttt{Replacement} is the traffic for DRAM cache replacement. 

%Similar to what we saw in \cref{sec:eval-tag}, 
Both Unison and Alloy Cache incur significant traffic for tag 
accesses. Alloy Cache also consumes considerable traffic for
speculative loads at cache misses. Unison Cache has small speculative 
load traffic due to its low miss rate. Both schemes also require 
significant replacement traffic.
Stochastic replacement can reduce Alloy Cache's
replacement traffic, but other overheads still remain. 

TDC can eliminate the tag traffic by managing mapping information in 
PTE/TLBs. However, like Unison Cache, it still incurs significant 
traffic for DRAM cache replacement. For most benchmarks, the traffic 
difference between Unison and TDC is just the removal of \texttt{Tag} 
traffic. For some benchmarks (e.g., \texttt{mcf}, 
\texttt{libquantum}), TDC incurs less replacement traffic than Unison 
Cache because of its higher hit rate due to full associativity. On 
some other benchmarks (e.g., \texttt{pagerank}, \texttt{tri\_count}), 
however, it incurs more traffic due to FIFO replacement. Overall, 
replacement traffic limits the performance of both Unison Cache and 
TDC.   

Because of the bandwidth-aware replacement policy, \name provides 
significantly better efficiency in in-package DRAM (35.8\% less 
traffic than the best baseline). 
%In fact, over 80\% of in-package DRAM traffic is serving cache hits 
%and the meta data traffic is less than 20\%. 
\name achieves this without incurring extra off-package traffic, which 
is a necessity to provide better performance.  On average, its 
off-package DRAM traffic is 3.1\% lower
than the best Alloy Cache scheme (\texttt{Alloy 1}), 42.4\% lower than 
Unison Cache and 43.2\% lower than TDC.

As mentioned earlier, graph codes are arguably more important for 
our modeled system.  We note that for graph codes with high traffic 
(i.e., \texttt{pagerank}, \texttt{tri\_count} and \texttt{graph500}), 
\name gives some of its largest gains, significantly reducing both in- 
and off-package DRAM traffic compared to all baseline schemes.
%from XYZ\% to ZYX\% over \texttt{Alloy 1}.

%\name incurs little traffic overhead compared to the other designs  
%and thus achieves better in-package DRAM efficiency. For most 
%benchmarks, \name can also reduce off-package DRAM traffic due to its 
%smarter replacement policy. For a few other benchmarks,
%%(e.g., \texttt{lbm} and \texttt{omnetpp})
%however, \name incurs more traffic due to its selective data caching 
%(\texttt{lbm}) or coarse granularity (\texttt{omnetpp}).  
\begin{figure}[t!]
    \centering
    \includegraphics[width=\columnwidth]{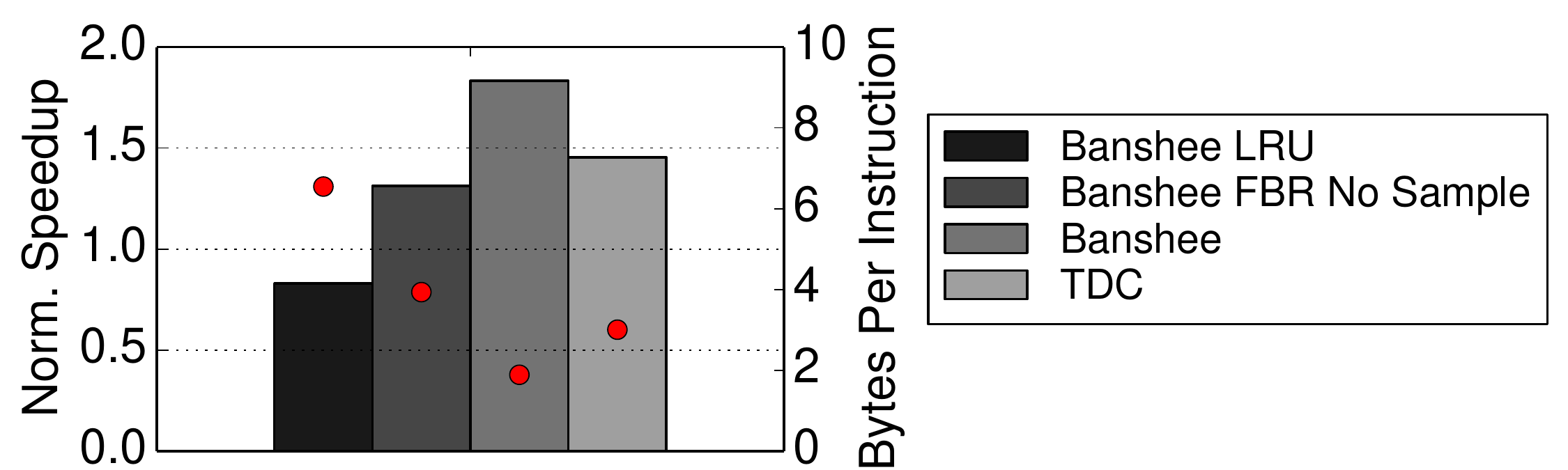}
    \vspace{-.2in}
    \caption{ \rev{Performance (bars) and DRAM cache bandwidth (red 
    dots) of different replacement policies on \mbox{\name} normalized 
    to \texttt{NoCache}. Results averaged over all benchmarks.} }
    \vspace{-.1in}
    \label{fig:banshee-variants}
\end{figure}

\subsection{\name Extensions} 

\subsubsection{Supporting Large Pages}

We evaluated the performance of large pages in \name for graph 
benchmarks. For simplicity, we assume all data resides on large (2~MB) 
pages. The sampling coefficient was chosen to be 0.001 and the 
replacement threshold was calculated accordingly 
(\cref{sec:repl-alg}).  

Our evaluation shows that with large pages, performance is on average 
3.6\% higher than the baseline Banshee with regular 4~KB pages. Here 
we assume perfect TLBs to only show the performance difference due to 
the DRAM subsystem. The gain comes from the more accurate hot page 
detection at larger page granularity as well as fewer frequency 
counter updates and PTE/TLB updates.  

\subsubsection{Balancing DRAM Bandwidth} \label{sec:balance}

Some related work~\cite{batman, agarwal2015unlocking, agarwal2015page} 
proposed to balance the accesses to in- and off-package DRAMs in order 
to maximize the overall bandwidth efficiency. These optimizations are 
orthogonal to \name and can be used on top of it.  

We implemented the technique from BATMAN~\cite{batman} which turns off 
parts of the in-package DRAM if it has too much traffic (i.e., over 
80\% of total DRAM traffic). On average, the optimization leads to 5\% 
(up to 24\%) performance improvement for Alloy Cache and 1\% (up to 
11\%) performance improvement for \name. The gain is smaller in \name 
since it has less total bandwidth consumption. With bandwidth 
balancing, \name still outperforms Alloy Cache by 12.4\%.  

\subsection{Sensitivity Study} \label{sec:eval-sensitivity} 

In this section, we study the performance of \name with different 
design parameters.  

\subsubsection[title]{ \rev{DRAM Cache Replacement Policy} }

\rev{We show performance and DRAM cache bandwidth of different 
replacement policies in \cref{fig:banshee-variants} to understand 
where the performance gain of \mbox{\name} comes from.} 

\texttt{Banshee LRU} uses an LRU policy similar to 
\texttt{UnisonCache} but does not use footprint cache. It has bad
performance and high bandwidth consumption due to its frequent page 
replacement (on every miss). 
%The performance is worse than TDC since we assumed perfect footprint 
%cache for TDC but always replace full pages in \texttt{Banshee LRU}.}

\rev{Using FBR improves performance and bandwidth efficiency on top of 
LRU since only hot pages are cached. However, if the frequency 
counters are updated on every DRAM cache access (\texttt{Banshee FBR 
no sample}, similar to CHOP~\cite{jiang2010}), significant meta data 
traffic ($2\times$ of \mbox{\name}) will be incurred leading to 
performance degradation.  We conclude that both FBR and sampling-based 
counter management should be used to achieve good performance in 
\mbox{\name}.}

\subsubsection{Page Table Update Overhead} \label{sec:eval-pt}

\begin{table}
    \caption{ Page table update overhead }
    \vspace{-0.1in}
    { \footnotesize
    \begin{tabular}{|c|c|c|}
    \hline
    Update Cost (us) & Avg Perf. Loss & Max Perf. Loss \\
    \hline
    10 & 0.11\% &  0.76\% \\
    \hline
    20 & 0.18\% & 1.3\% \\
    \hline
    40 & 0.31\% & 2.4\% \\
    \hline
    \end{tabular}
    }
    \label{tab:coherence}
\end{table}

One potential disadvantage of \name is the overhead of page table 
updates (cf.  \cref{sec:coherence}). However, this cost is paid only 
when the tag buffer fills up after many page remappings.
Furthermore, our replacement policy intentionally slows remapping
(cf. \cref{sec:fbr}). On average, the page table update is triggered   
once every 14 milliseconds, which has low overhead in practice. 

\cref{tab:coherence} shows the average and maximal performance 
degradation across our benchmarks, relative to free updates, for
a range of update costs.
The average overhead is less than 1\%, and scales sublinearly with
update cost. Note that doubling the tag buffer size has similar effect 
as reducing the page table update cost by half. Therefore, we do not 
study the sensitivity of tag buffer size here.  

\subsubsection{DRAM Cache Latency and Bandwidth}

\begin{figure}[]
	\centering
	\subfloat{\includegraphics[width=.9\columnwidth]{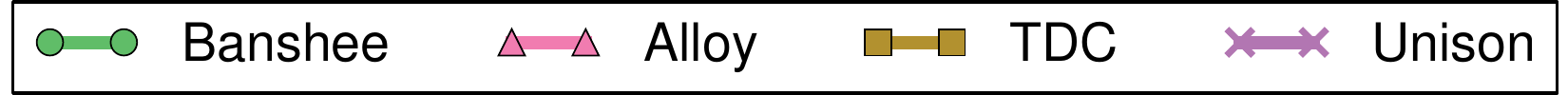}} 
	\\
	\vspace{-0.14in}
	%\vspace{-0.15in}
	\subfloat[Latency]{
		\includegraphics[width=.45\columnwidth]{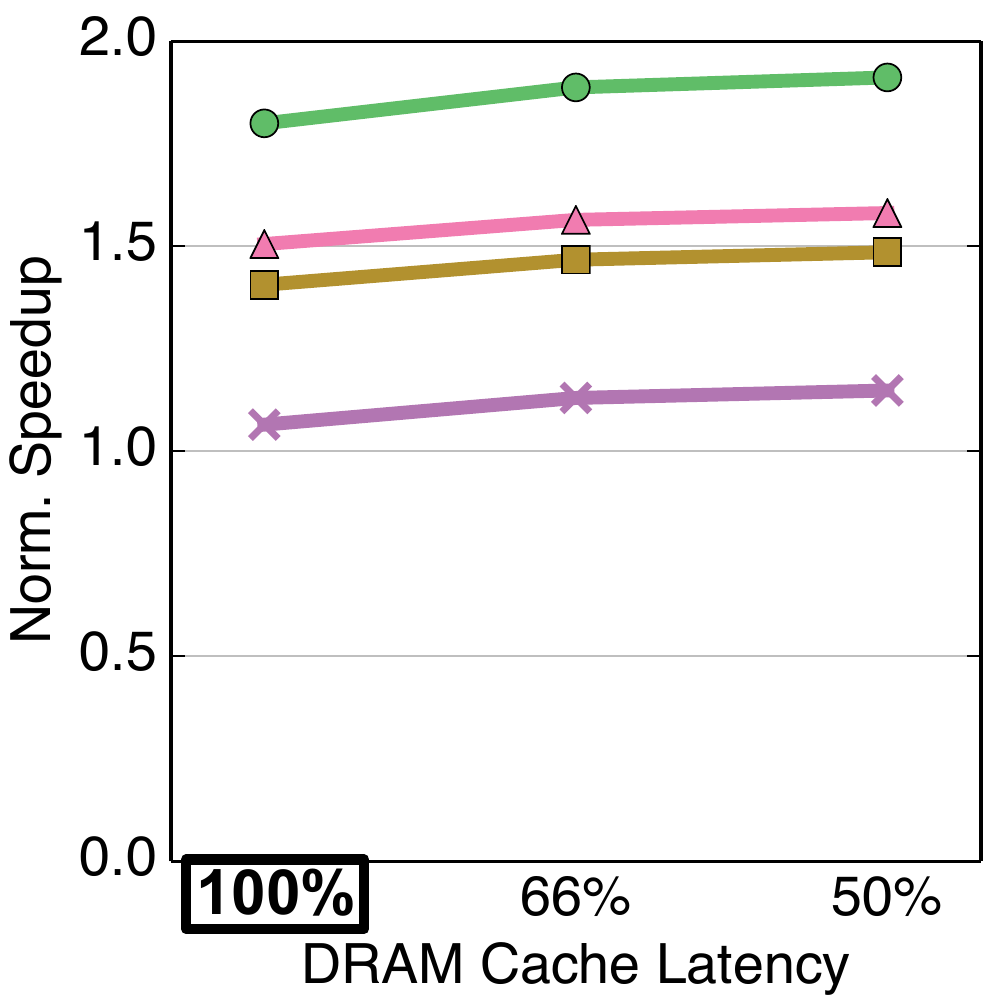}
		\label{fig:latency}
	} %\vspace{-0.15in}
	\hfill
    \subfloat[Bandwidth]{
		\includegraphics[width=.45\columnwidth]{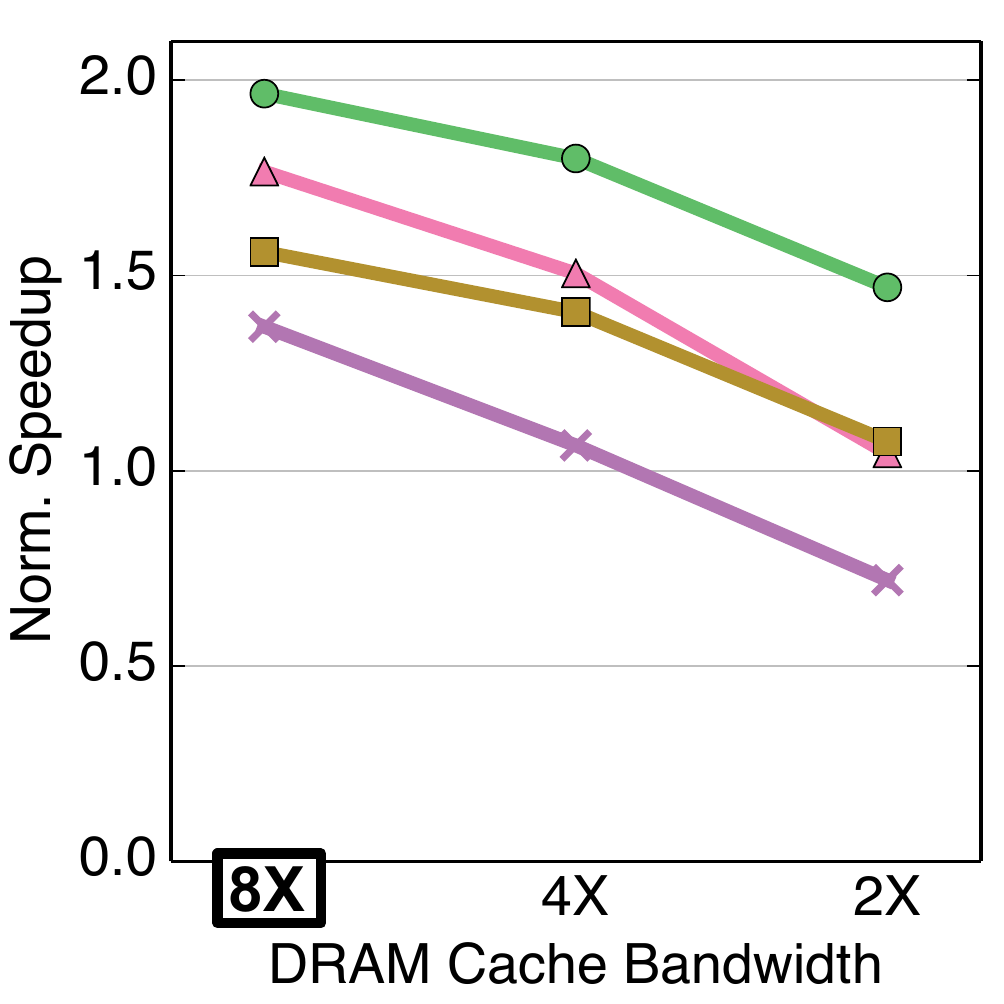}
        \label{fig:bandwidth}
    }
    \vspace{-0.1in}
    \caption{ Sweeping DRAM cache latency and bandwidth. Default 
    parameter setting highlighed on x-axis.}
    \vspace{-0.1in}
    \label{fig:lat-bw}
\end{figure}

\cref{fig:lat-bw} shows the performance (normalized to 
\texttt{NoCache}) of different DRAM cache schemes sweeping the DRAM 
cache latency and bandwidth. Each data point is the geometric mean 
performance over all benchmarks. The x-axis of each figure shows the 
latency and bandwidth of in-package DRAM relative to off-package DRAM.  
By default, we assume in-package DRAM has the same latency and 
$4\times$ bandwidth as off-package DRAM. 

As the in-package DRAM's latency decreases and bandwidth increases, 
performance of all DRAM cache schemes gets better. We observe that
performance is more sensitive to bandwidth change than to zero-load 
latency change, since bandwidth is the bottleneck in these workloads.  
Although not shown in the figure, changing the core count in the 
system has a similar effect as changing DRAM cache bandwidth. Since  
\name's performance gain over baselines is more significant when the 
bandwidth is more limited, we expect \name to perform better with more 
cores.

\subsubsection{Sampling Coefficient}

\begin{figure}[]
    \centering
    \subfloat[Miss rate]{
        \includegraphics[width=.45\columnwidth]{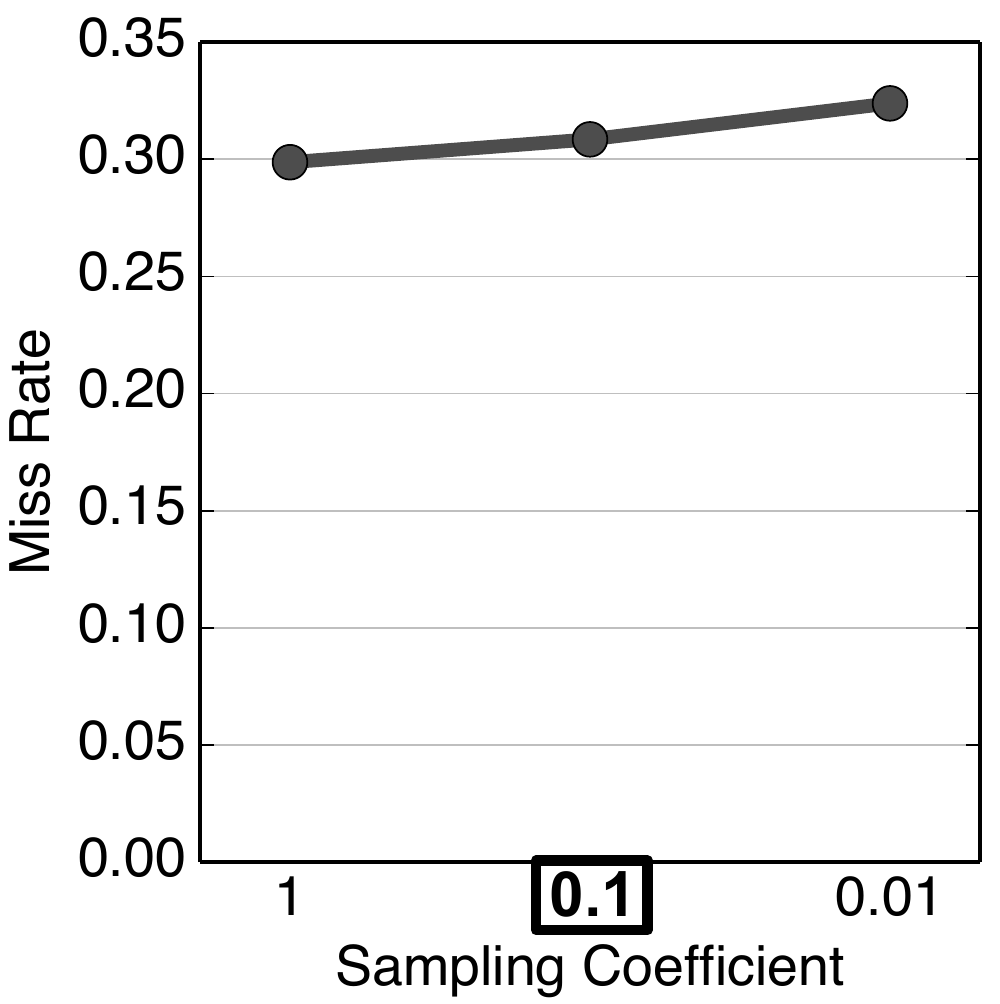}
        \label{fig:miss-sample}
    } %\vspace{-0.15in}
	\hfill
    \subfloat[DRAM cache traffic]{
        \includegraphics[width=.45\columnwidth]{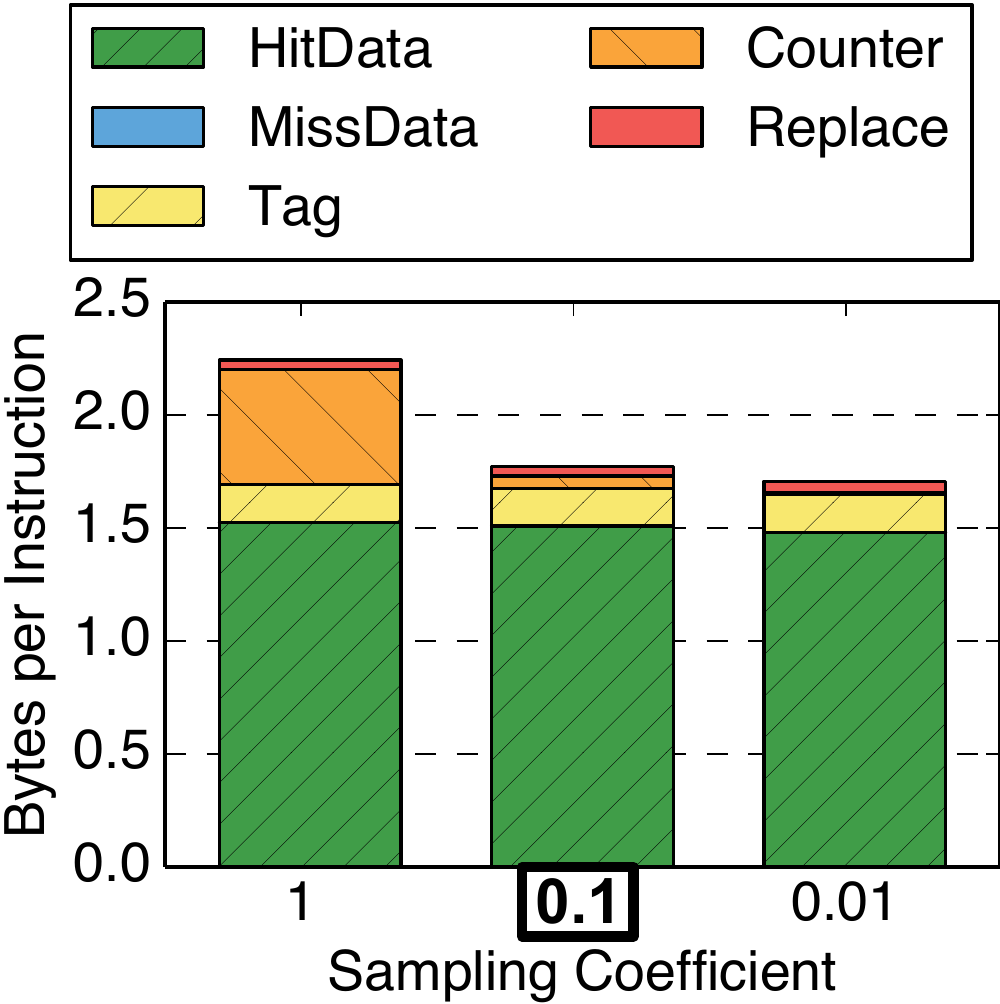}
        \label{fig:mc-sample}
    } \vspace{-.1in}
    \caption{ Sweeping sampling coefficient in \name (default sampling 
    coefficient = 0.1).  }
    \vspace{-.1in}
    \label{fig:sample}
\end{figure}

\cref{fig:sample} shows the DRAM cache miss rate and traffic breakdown 
sweeping the \textit{sampling coefficient} in \name. As the sampling 
coefficient decreases, miss rate increases but only by a small amount.

\name incurs some traffic for loading and updating frequency counters, 
but this overhead becomes negligible for a sampling rate of $\le$10\%,
which still provides a low miss rate.

\subsubsection{Associativity}

\begin{table}[!t]
    \caption{ Cache miss rate vs. associativity in \name }
	\vspace{-.05in}
    { \footnotesize
    \begin{tabular}{|c||c|c|c|c|}
    \hline
    Number of Ways & 1 way & 2 ways & 4 ways & 8 ways \\
    \hline
    Miss Rate & 36.1\% & 32.5\% & 30.9\% & 30.7\% \\
    \hline
    \end{tabular}
    }
	\vspace{-.05in}
    \label{tab:missrate-way}
\end{table}

%\begin{figure}[t!]
%    \centering
%    \includegraphics[width=.9\columnwidth]{figs/missrate_way.pdf}
%    \caption{ Cache miss rate vs. associativity in \name.}
%    \label{fig:missrate-way}
%\end{figure}

In \cref{tab:missrate-way}, we sweep the number of ways in \name and 
show the cache miss rate.
Doubling the number of ways requires adding one more 
bit to each PTE, and doubles the per-set metadata.
Higher associativity reduces the cache miss rate, though.
Since we see quickly diminishing gains above four ways,
we choose that as the default design point.

\section{Related Work}

Besides those discussed in \cref{sec:back}, other DRAM cache designs 
have been proposed in the literature. PoM~\cite{sim2014} and 
CAMEO~\cite{chou2014} manage in- and off-package DRAM in different
address spaces at cacheline granularity.  Tag Tables~\cite{franey2015} 
compressed the tag storage for Alloy Cache
so that it fits in on-chip SRAM. Bi-Modal Cache~\cite{gulur2014} 
supports heterogeneous block sizes (cacheline and page) to get the 
best of both worlds. All these schemes focus on minimizing latency of 
the design and incur significant traffic for tag lookups and/or cache 
replacement. 

Similar to this paper, several other papers have proposed DRAM cache 
designs with bandwidth optimizations.  CHOP~\cite{jiang2010} targets 
the off-package DRAM bandwidth bottleneck for page granularity DRAM 
caches, and uses frequency-based replacement instead of LRU.  However, 
their scheme still incurs significant traffic for counter updates, 
whereas \name uses sampling based counter management and
bandwidth-aware replacement.  Several other papers propose to 
improve off-package DRAM traffic for page granularity
using a \textit{footprint 
cache}~\cite{jevdjic2013, jevdjic2014, jang2016}. As we showed,
however, this alone cannot eliminate all unnecessary replacement 
traffic. That said, the footprint idea is orthogonal to \name and 
therefore can be incorporated to \name for even better performance.  
%to predicting which blocks on a
%page will be used can provide a significant traffic reduction ---
%we should be able to combine this idea with \name.

BEAR~\cite{chou2015} improves Alloy Cache's DRAM cache bandwidth
efficiency.  Our implementation of Alloy Cache already 
includes some of the BEAR optimizations. These optimizations 
cannot eliminate all tag lookups, and as we have shown in 
\cref{sec:eval-traffic}, \name provides higher DRAM cache 
bandwidth efficiency.

Several other works have considered heterogeneous memory
technologies beyond in--package DRAM.
These include designs for hybrid 
DRAM and Phase Change Memory (PCM)~\cite{meza2012, yoon2012, 
dhiman2009} and a single DRAM chip with fast and slow 
portions~\cite{lee2013, lu2015}.  We believe the ideas in this paper 
can be applied to such heterogeneous memory systems, as well.  

\rev{Among all previous designs, TDC~\cite{lee2015} is the one closest
to \mbox{\name}. Both schemes use PTE/TLBs to track data mapping at 
page granularity. The key innovation in \mbox{\name} was the 
bandwidth-efficient DRAM cache replacement policy and the associated 
designs that enabled it (lazy TLB coherence). \mbox{\name} 
significantly reduces both data and meta data replacement traffic 
through FBR and frequency counter sampling. This improves in- and 
off-package DRAM bandwidth efficiency which leads to performance 
improvement.}

The replacement policy used in \name, however, cannot be efficiently 
implemented on TDC due to the address consistency and TLB coherence 
problem. Since TDC uses different physical addresses for in- and 
off-package DRAMs, if a page is remapped after some of its cachelines 
have been caches, these previously loaded cachelines will have stale 
addresses.  This makes the existing address consistency problem in TDC 
even worse.  
%Further, TLB coherence in TDC is achieved through the global TLB 
%directory.  When a page is remapped after its first miss, stale 
%TLBs/PTEs need to be invalidated/updated which incurs complexity and 
%performance loss.  Note that this was not a problem in the original 
%TDC algorithm which does replacement on the first miss when no TLB 
%has an record for the page and thus no invalidation is needed. 

%\subsection{Cache Replacement Policy}
%
%Cache replacement policy has been studied by the community for 
%decades. On chip caches often use LRU or its variants and numerous 
%optimizations have been proposed~\cite{qureshi2007, jaleel2010, ??, 
%??, ??}. 
%
%However, DRAM cache has quite different design requirement as 
%traditional SRAM caches. Due to architectural differences, reading and 
%updating  replacement metadata can be very expensive in DRAM while it 
%is cheap in SRAM. Cache replacement itself also becomes more expensive 
%in a DRAM cache. Therefore, the optimizations proposed for traditional 
%SRAM  caches may not be directly applicable to DRAM caches.
%
%Similar to this paper, several previous works proposed to use 
%frequency based policies to make smarter replacement decision.  
%However, they either require storing the frequency counters on 
%chip~\cite{??, ??} or require updating the counters in DRAM on every 
%cache miss~\cite{??}. As a result,  
%
%\TODO{Give examples of other frequency based schemes. }
%
%\TODO{Talk about filtering (sampling) in cache replacement. }
%

\section{Conclusion}

A new DRAM cache algorithm, \name, was proposed in this paper. \name 
aims at maximizing in- and off-package DRAM bandwidth efficiency and 
therefore performs better than previous latency optimized DRAM cache 
algorithms. \name achieves this through a software hardware co-design 
approach. Specifically, \name uses a new TLB coherence mechanism, and 
a bandwidth aware DRAM replacement policy. Our extensive experimental 
results show that \name can provide significant improvement over 
state-of-the-art DRAM cache schemes.

%For systems with in-package DRAM that provides bandwidth rather
%than latency benefits, the DRAM cache should be designed with those
%characteristics in mind.
%Recent designs on DRAM caches optimized for hit latency
%incurs extra DRAM traffic for meta data management and cache 
%replacement, leading to suboptimal performance. by leveraging 
%presence and way information in the page table, using a software 
%hardware co-designed TLB coherence mechanism, and having a bandwidth 
%aware DRAM replacement policy. Our extensive experimental results 
%show that \name can provide significant improvement over 
%state-of-the-art DRAM cache schemes.   

%\newpage
{
	%\bstctlcite{bstctl:etal, bstctl:nodash, bstctl:simpurl}
    \bibliographystyle{ieeetr}
    \bibliography{refs}
}
%\newpage
%\appendix
%\input{appendix}

\end{document}